\documentclass{aastex63}
\usepackage{booktabs}
\usepackage{graphicx}
\usepackage{wrapfig}
\usepackage{lineno}
\DeclareUnicodeCharacter{2212}{-}

\makeatletter
\def\@to{to}
\makeatother

\received{\today}
\revised{XXX}
\accepted{XXX}
\submitjournal{AJ}

\shorttitle{Methanol maser study of G358.93$-$0.03-MM1}
\shortauthors{Bayandina et al.}
\graphicspath{{./}{figures/}}

\begin{document}

\title{A multi-transition methanol maser study of the accretion burst source G358.93$-$0.03-MM1}

\correspondingauthor{Olga Bayandina}
\email{bayandina@jive.eu}

\author[0000-0003-4116-4426]{O. S. Bayandina}
\affiliation{Joint Institute for VLBI ERIC, Oude Hoogeveensedijk 4, 7991 PD Dwingeloo, The Netherlands}
\affiliation{Astro Space Center, P.N. Lebedev Physical Institute of RAS, 84/32 Profsoyuznaya st., Moscow, 117997, Russia}

\author[0000-0002-6558-7653]{C. L. Brogan}
\affiliation{National Radio Astronomy Observatory, 520 Edgemont Road, Charlottesville, VA 22903, USA}

\author[0000-0003-3302-1935]{R. A. Burns}
\affiliation{Joint Institute for VLBI ERIC, Oude Hoogeveensedijk 4, 7991 PD Dwingeloo, The Netherlands}
\affiliation{Mizusawa VLBI Observatory, National Astronomical Observatory of Japan, 2-21-1 Osawa, Mitaka, Tokyo 181-8588, Japan}

\author{X. Chen}
\affiliation{Center for Astrophysics, Guangzhou University, Guangzhou 510006, People's Republic of China}
\affiliation{Shanghai Astronomical Observatory, Chinese Academy of Sciences, 80 Nandan Road, Shanghai, 200030, People's Republic of China}

\author[0000-0001-6492-0090]{T. R. Hunter}
\affiliation{National Radio Astronomy Observatory, 520 Edgemont Road, Charlottesville, VA 22903, USA}

\author[0000-0003-4444-5602]{S.~E. Kurtz}
\affiliation{Instituto de Radioastronom{\'\i}a y Astrof{\'\i}sica, Universidad Nacional Aut\'onoma de M\'exico, Apdo. Postal 3-72, Morelia, 58089, Mexico}

\author{G. C. MacLeod}
\affiliation{The Open University of Tanzania, P.O. Box 23409, Dar-Es-Salaam, Tanzania}
\affiliation{Hartebeesthoek Radio Astronomy Observatory, PO Box 443, Krugersdorp, 1741, South Africa}

\author[0000-0002-3773-7116]{A. M. Sobolev}
\affiliation{Ural Federal University, 51 Lenin Str., 620051 Ekaterinburg, Russia}

\author[0000-0002-6033-5000]{K. Sugiyama}
\affiliation{Mizusawa VLBI Observatory, National Astronomical Observatory of Japan, 2-21-1 Osawa, Mitaka, Tokyo 181-8588, Japan}

\author[0000-0003-4032-5590]{I. E. Val'tts}
\affiliation{Astro Space Center, P.N. Lebedev Physical Institute of RAS, 84/32 Profsoyuznaya st., Moscow, 117997, Russia}

\author[0000-0001-5615-5464]{Y. Yonekura}
\affiliation{Center for Astronomy, Ibaraki University, 2-1-1 Bunkyo, Mito, Ibaraki 310-8512, Japan}



\begin{abstract}
We present the most complete to date interferometric study of the centimeter wavelength methanol masers detected in G358.93$-$0.03 at the burst and post-burst epochs. A unique, NIR/(sub)mm-dark and FIR-loud MYSO accretion burst was recently discovered in G358.93$-$0.03. The event was accompanied by flares of an unprecedented number of rare methanol maser transitions. The first images of three of the newly-discovered methanol masers at  6.18,  12.23,   and  20.97  GHz are presented in this work. The spatial structure evolution of the methanol masers at 6.67, 12.18, and 23.12 GHz is studied at two epochs. The maser emission in all detected transitions resides in a region of $\sim$0.2$^{\prime\prime}$ around the bursting source and shows a clear velocity gradient in the north-south direction, with red-shifted features to the north and blue-shifted features to the south. 
A drastic change in the spatial morphology of the masing region is found: a dense and compact "spiral" cluster detected at epoch I evolved into a disperse, "round" structure at epoch II. 
During the transition from the first epoch to the second, the region traced by masers expanded.
The comparison of our results with the complementary VLA, VLBI, SMA, and ALMA maser data is conducted. The obtained methanol maser data support the hypothesis of the presence of spiral-arm structures within the accretion disk, which was suggested in previous studies of the source.

\end{abstract}

\keywords{masers --- stars: individual (G358.93−0.03) --- stars: formation}


\section{Introduction} \label{sec:intro}

Is there a common mechanism of star formation across the entire stellar mass spectrum? 
While star formation theory is well established for low-mass stars (e.g., \citealt{McKee07}), much less is known about high-mass star formation. 
Recent studies such as \cite{Caratti17} and \cite{Hunter17} provide direct evidence that massive young stellar objects (MYSOs), analogously to low-mass stellar objects, form via disk-mediated accretion accompanied by episodic accretion bursts.  These bursts may be a result of disk fragmentation \citep{Ahmadi19, Meyer19}. 
Such a mechanism provides a means to overcome radiation pressure \citep{Hosokawa16}, and it is thought that up to half of the final stellar mass may be acquired in these accretion events. 
But the accretion process itself is poorly understood --- largely due to scarce observational evidence. 
Presently, only three cases of accretion bursts in MYSOs have been reported: S255IR \citep{Caratti17}, NGC6334I \citep{Hunter17}, and --- the topic of this paper --- G358.93$-$0.03 \citep{stecklum21}. A potential fourth accretion-burst source, G323.46$-$0.08, may have recently undergone such an event \citep{Proven19}.

A larger sample is clearly needed, but identifying MYSOs undergoing accretion bursts is difficult.
It is in this respect that masers have proven to be a powerful probe of the mechanisms of massive star formation. Masing occurs only within certain ranges of physical conditions of the gas and radiation field (e.g., see the reviews by \citealt{Ellingsen07} and \citealt{Breen19}). Thus the spatial distribution of masers can reveal the distribution of temperature, density and radiation enhancements in the region, while the kinematics of maser spots can trace gas motions. All of these properties are essential for understanding the episodic accretion phenomenon.

If the accretion burst augments the local radiation field, the resulting  increase in incident photons will cause all masers in the foreground of the continuum emission to increase in flux. Masers, either compact or extended, and covering a wide range of velocities may be involved --- as was seen in S255IR \citep{Szymczak18}, NGC6334I \citep{MacLeod18}, and recently in G358.93$-$0.03 \citep{MacLeod19, Brogan19, Breen19}. Moreover, in all these sources, multiple maser transitions were seen to flare\footnote{Hereafter, we use the word "flare" to refer to a sudden increase in the maser flux density (in some cases, caused by an accretion burst) and the word "burst" to refer to the accretion burst itself.} and rare maser transitions appeared.  The masers then weakened or disappeared in a span of weeks to months for the rarer masers but longer for the more common transitions \citep{MacLeod18}.

\subsection{G358.93-0.03}
\begin{figure*}[ht]
\centering
\gridline{\fig{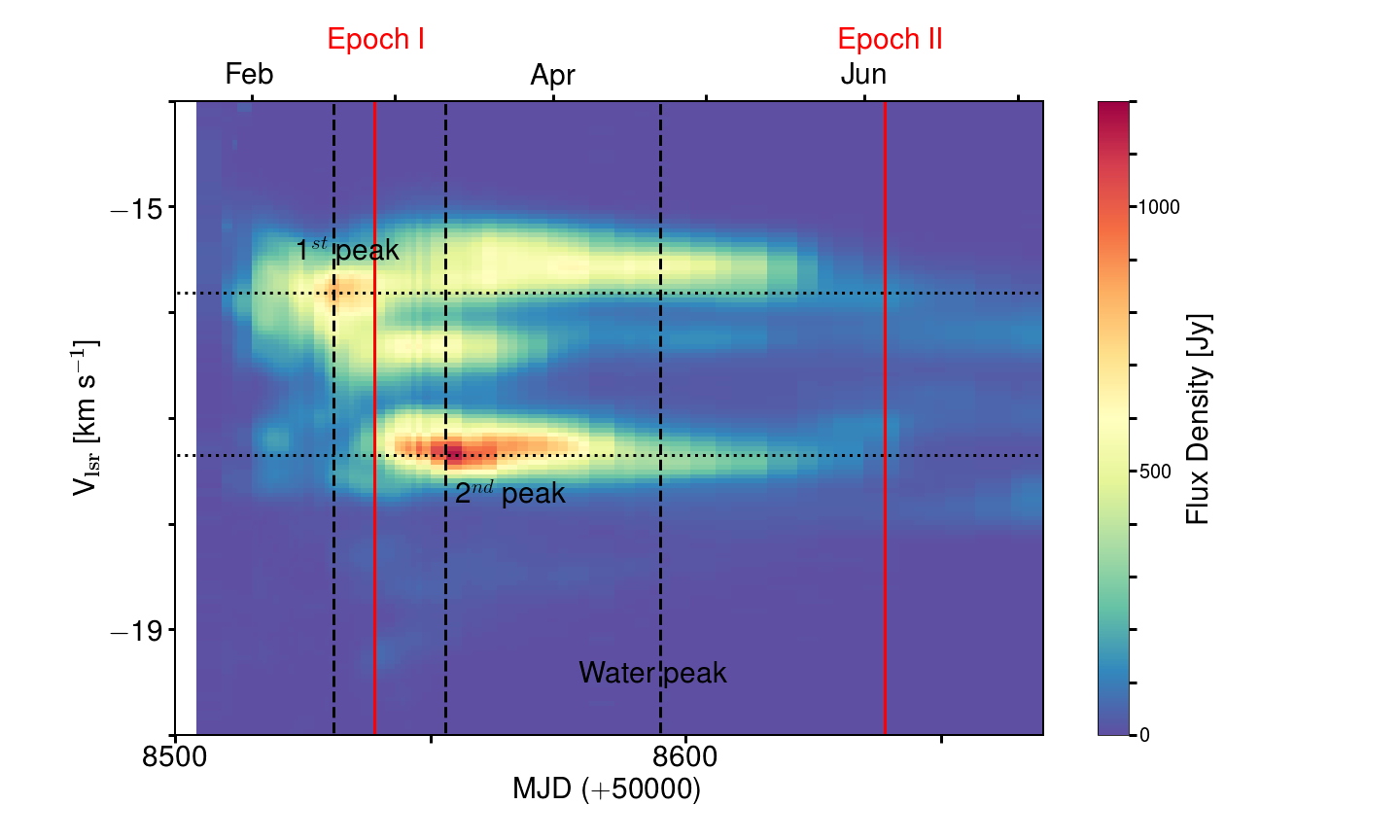}{0.6\textwidth}{} }
\caption{The dynamic spectrum of the 6.67 GHz methanol 
maser emission in G358.93$-$0.03 for the period January--July 2019. The black dotted lines indicate flares of the methanol and water masers. The red lines indicate the dates of the VLA epoch I and II observations (M2O data: the HartRAO 26-m telescope monitoring program).
\label{fig:timeline}}
\end{figure*}

On January 14, 2019, a rapid rise of the 6.67 GHz class II methanol maser flux density, accompanied by the appearance of several new velocity features, was detected in G358.93$-$0.03 \citep{Sugiyama19} by the Ibaraki 6.7 GHz Methanol Maser Monitor (iMet) program \citep{Yonekura2016}. The flare of 6.67 GHz maser emission reached a first peak on  February 15, 2019, when the flux density of the maser feature at $-15.6$ km~s$^{-1}$ reached $\sim$660~Jy (Figure~\ref{fig:timeline}). A second peak occurred on March 12, when the velocity feature at $-17.2$ km~s$^{-1}$ reached a flux density of $\sim$900~Jy.
  A daily spectrum of the 6.67~GHz methanol maser in G358.93$-$0.03 is published regularly on the iMet project website\footnote{\url{http://vlbi.sci.ibaraki.ac.jp/iMet/G358.9-00-190114/daily.html}}.
  
Intensive follow-up observations were made by the Maser Monitoring Organization\footnote{\url{https://www.masermonitoring.org/}} (M2O), a self-organized collaboration of maser monitoring observatories.
Monitoring with the 26-m telescope of the Hartebeesthoek Radio Astronomy Observatory (HartRAO, Republic of South Africa; Figure~\ref{fig:timeline}) confirmed that the flaring 6.67 GHz maser flux density had been rapidly increasing since February of 2019
and in $\sim$1.5 months reached $\sim$1000 Jy --- about 200$\times$ greater than the stable flux density of $\sim$5 Jy that had been reported by observations spanning the previous decade \citep{Caswell10, Chambers14, Hu16, Rickert19}.

In addition to the 6.67 GHz methanol maser flare, the appearance of multiple new maser transitions was observed in G358.93$-$0.03.  A rare 23.12 GHz class II methanol maser (only the fourth known occurrence of this transition) was detected. A 12.18 GHz class II methanol maser, previously undetected in this source despite several epochs of observations \citep{Breen12}, was found to have a higher flux density than the flaring 6.67 GHz maser. 
The ATCA  observations  of  G358.93$-$0.03 conducted in March, 2019 detected  emission  from  12  additional  methanol  transitions,  six  of which  were  previously  undetected  class  II  methanol  masers;  among these were the  first known torsionally  excited  methanol  masers  \citep{Breen19}. In January-March, 2019, \cite{Volvach20} reported a short-lived (44 days) 19.97 GHz methanol maser emission. More transitions may have been missed owing to insufficient frequency coverage.  The latest observations  with  the SMA (March, 2019) and ALMA (April, 2019) yielded an additional 14 methanol maser discoveries in the (sub)millimeter range, primarily from transitions in excited torsional states \citep{Brogan19}. 

G358.93$-$0.03  was  a  maser-quiet  source  at the  pre-flare epoch, with only two maser transitions detected: the 6.67~GHz methanol maser \citep{Caswell10, Chambers14, Hu16, Rickert19} and 22 GHz water maser \citep{Titmarsh16}.  Moreover, G358.93$-$0.03 was not as well-studied as S255IR or NGC6334I. For example, no maser proper motion data is available to provide a parallax distance. According to the BeSSeL Revised Kinematic Distance Calculator\footnote{\url{http://bessel.vlbi-astrometry.org/revised_kd_2014}} \citep{Reid2014}, the near kinematic distance to the source is $6.75^{+0.37}_{-0.68}$~kpc.

The SMA and ALMA (sub)millimeter imaging resolved G358.93$-$0.03 into a cluster of eight continuum sources \citep{Brogan19}. Two of these sources, MM1 and MM3, were found to host hot molecular cores. 
All of the newly-discovered methanol masers were found to be associated with the brightest continuum source, MM1 (e.g. \cite{Brogan19, Burns20}).

The flaring class II methanol maser transitions detected in G358.93$-$0.03 require similar physical conditions according to theoretical models, albeit over a wider range of values \citep{Sobolev97a,Sobolev97b,Cragg04,Cragg05}.  Thus, the simultaneous flaring of these masers indicates a sudden change in the local physical conditions --- probably related to the radiation field --- which could provoke the necessary conditions for maser amplification.  

The multi-epoch Long Baseline Array (LBA) observations, performed during the 6.67 GHz methanol maser flare, allowed the imaging of the subluminal propagation of a thermal radiation "heatwave" emanating from the accreting high-mass protostar \citep{Burns20}. 

The SMA and ALMA images revealed a partial elliptical ring of the methanol masers, which was interpreted as a coherent physical structure illuminated by a radiative event from the central object \citep{Brogan19}.

Remarkably, the VLA observations of \cite{Chen20a,Chen20b}, conducted during the maser flare epoch in March, 2019, revealed that the discovered rare maser transitions of HDO, HNCO, and $^{13}$CH$_3$OH appear to trace a spiral-arm structure around the bursting source \citep{Chen20b}. This finding was the first observational evidence of a link between accretion bursts and disk substructures \citep{Chen20b}.

The accretion burst in G358.93$-$0.03 was decisively confirmed by multi-epoch SOFIA observations \citep{stecklum21}. The event is found to be the first NIR/(sub)mm-dark and FIR-loud MYSO accretion burst, showing an increase in the source flux only in the FIR, and not in the NIR or (sub)mm bands \citep{stecklum21}.

The excellent coordination of the M2O observatories ensured that monitoring of G358.93$-$0.03 was done in a timely manner.  As a result, we had the unique opportunity to observe a stage of the flare that was missed in the cases of S255 and NGC6334I. In this paper, we present the most complete to date interferometric study of the methanol masers  detected in G358.93$-$0.03 at the burst and post-burst epochs.

\newpage
\section{OBSERVATIONS AND DATA REDUCTION \label{sec:obs}}

Two observing sessions of G358.93$-$0.03 were carried out with the Karl G. Jansky Very Large Array (VLA) on February 25, 2019 (epoch I) and on June 4, 2019 (epoch II)  as the Target of Opportunity programs 19A-448 and 19A-476. 
The first session was a follow-up observation in response to the rapid rise of the 6.67 GHz methanol maser flux density detected in the source in January 2019 \citep{Sugiyama19}. The second session was made in response to the 22 GHz water maser flare of April 2019 (the M2O monitoring data, Fig. \ref{fig:timeline}). 

The epoch I  observation was made during a C$\,\to\,$B reconfiguration which led to an asymmetrical beam, highly elongated in the N-S direction. At the  first epoch, the priority was to catch the flaring methanol masers before they faded, so promptness was the driving factor more than resolution. The epoch II observation was made in A-configuration. 

In this paper, we present the C-, Ku-, and K-bands spectral line and K-band continuum observations.
The observing time for each session was two hours.  The pointing coordinates for G358.93$-$0.03 were RA (J2000) = 17$^{\rm h}$ 43$^{\rm m}$ 10.02$^{\rm s}$ and Dec (J2000) = $-29^{\circ}$51$\arcmin$45.8$\arcsec$, with an 
LSR velocity of $-15.55$~km~s$^{-1}$.
The same calibration sources were used at both epochs: 3C286 was used as a flux density, bandpass, and delay calibrator; J1744$-$3116 was the complex gain calibrator (with an angular separation from the target source of 4.6$^{\circ}$).

The maser lines were observed in narrow spectral windows (1, 2 and 4 MHz at C-, Ku-, and K-bands, respectively) with 512 or 1024 channels. 
Continuum emission was observed in 31 spectral windows with 128 1-MHz channels. 
Observation parameters, including the synthesized beam size and the rms noise level, for spectral line and continuum data, are presented in Tables \ref{tab:obsspl} and \ref{tab:obsc}. Table \ref{tab:obsspl} contains the list of detected maser transitions (detection of maser emission in each band and epoch is marked with 'Y', non-detection is marked with 'N').

All steps of the post-correlation data reduction were done with the Common Astronomy Software Applications (CASA, \cite{McMullin07}). For basic flagging and data calibration, we used the VLA CASA Calibration Pipeline. Nevertheless, there were two issues that required special, additional treatment. 
First, during the observations we used the VLA catalog coordinates for the phase calibrator J1744$-$3116.  Post-observation, we learned that the VLBI Source Position Catalog\footnote{\url{http://astrogeo.org/vlbi/solutions/rfc\_2019d/}} lists a more reliable position.  The discrepancy from the VLA catalog position is 0.313 arcsec. 
Prior to calibration and imaging we corrected the calibrator position to the VLBI catalog values with CASA task {\it fixvis}.
Second, the observations were carried out during a multi-frequency maser flare in the source, which led to the detection of a number of methanol maser lines in the wide-band VLA continuum window (for the list of the detected maser lines, see the note to Table \ref{tab:obsc}). The maser lines were not flagged by the pipeline, and their data appeared in the continuum images as a false detection of continuum peaks. Additional manual flagging was done to avoid this contamination.

Calibrated data were imaged with the CASA task \textit{clean}. Briggs weighting was used for maser data and natural weighting was used for continuum data. Gaussian fitting of the images was performed with the CASA task \textit{imfit}. A two-dimensional Gaussian brightness distribution was fit to all  maser and continuum emission peaks with a flux density above the 3$\sigma$ level (see Tables \ref{tab:obsspl} and \ref{tab:obsc} for $1\sigma$-levels) to determine their positions and flux densities. 
Further in the text, we list and discuss so-called ``maser spots'' --- maser emission detected in a single velocity channel of a data cube.

To estimate the uncertainties in the absolute positions of the masers detected with the VLA, we compared our positions with those obtained by other observations made recently by the M2O collaboration. The 6.7 GHz methanol maser spots detected with the VLA at the first epoch (February 25, 2019) were superposed with the 6.7 GHz data obtained with the LBA on February 28, 2019 from \cite{Burns20}. A N-S position difference was noted and estimated to be $\sim$0.07 arcsec. This value is within the accuracy of the VLA position measurement and is within one pixel of our C-band VLA images. 
Absolute coordinates measured with VLA can be affected by uncompensated ionospheric propagation delays, and the effect is most notorious for declinations $\leq$30$^{\circ}$ \citep{ARM2000}. To estimate the displacements for other types of masers, we (1) measured the median positions of the centres of the maser clusters, (2) estimated shifts between centres of the maser clusters and MM1 (the central source of our maps \citep{Brogan19}). The estimated offsets are indicated in the notes of Tables \ref{tab:T67GHZ}-\ref{tab:T23GHZ}. The offsets were introduced to our data and used in the data analysis and preparation of Figures \ref{fig:67GHZ}-\ref{fig:23GHZ}, \ref{fig:compare2}-\ref{fig:overplot}, and \ref{fig:lba}-\ref{fig:comparechen}. 
Note that the absolute positional uncertainty of MM1, with which we compare positions of the detected masers, is $\sim$0.03 arcsec \citep{Brogan19}. The absolute position shift between the first and second VLA epochs appeared to be of $\sim$0.04 arcsec.

\begin{table}[ht]
\caption{Observation parameters: Spectral Lines}
\label{tab:obsspl}
\resizebox{\textwidth}{!}{
\begin{tabular}{cccccccccc}
\hline \hline
Band & Transition & Freq.\tablenotemark{\footnotesize{a}} & Epoch & Int. Time &
Synth. Beam & PA & Spec. res. & $1\sigma$~rms & Detect.\\ 
 &  & (MHz) & & (min) &
(arcsec) & ($^\circ$) & (km~s$^{-1}$) & (mJy/beam) & \\ \hline
C  & CH$_3$OH 17$_{-2}$ $\rightarrow$ 18$_{-3}$ E (v$_t$=1) & 6181.146(21)\tablenotemark{\footnotesize{1}} & II & 16 & 2.02~$\times$~0.86 & +0.12 & 0.09 & 10 & Y \\
& CH$_3$OH 5$_{1}$ $\rightarrow$ 6$_0$ A$^+$ (v$_t$=0) & 6668.5192(8)\tablenotemark{\footnotesize{2}} & I & 14 & 3.65~$\times$~0.75 & +7.44 & 0.09 & 12 & Y \\
  &  &   & II & 16 & 1.89~$\times$~0.80 & +0.41 & 0.09 & 8 & Y\\
   \hline
Ku & CH$_3$OH 2$_0$ $\rightarrow$ 3$_{-1}$ E (v$_t$=0) & 12 178.597(4)\tablenotemark{\footnotesize{2}} & I & 16 & 1.97~$\times$~0.40 & +4.36 & 0.10 & 10 & Y \\
   &   &   & II & 16 & 1.10~$\times$~0.43 & -5.16 & 0.10 & 6 & Y \\
   & CH$_3$OH 16$_5$ $\rightarrow$ 17$_4$ E (v$_t$=0) & 12 229.348(16)\tablenotemark{\footnotesize{1}} & II & 16 & 1.08~$\times$~0.43 & -4.95 & 0.10 & 7 & Y \\
   \hline
K  & CH$_3$OH 17$_{6}$ $\rightarrow$ 18$_{5}$ E (v$_t$=0) &  20 346.864(16)\tablenotemark{\footnotesize{1}} & II & 19 & 0.67~$\times$~0.25 & +10.85 & 0.12 & 10 & N \\
   & CH$_3$OH 10$_{1}$ $\rightarrow$ 11$_{2}$ A$^+$ (v$_t$=1) & 20 970.620(21)\tablenotemark{\footnotesize{1}} & II & 19 & 0.65~$\times$~0.24 & -10.49 & 0.11 & 11 & Y \\
  & CH$_3$OH 9$_{2}$ $\rightarrow$ 10$_{1}$ A$^+$ (v$_t$=0) & 23 121.0242(5)\tablenotemark{\footnotesize{2}} & I & 16 & 1.00~$\times$~0.20 & -1.18 & 0.05 & 15 & Y \\
   &   &   & II & 19 & 0.58~$\times$~0.22 & -11.28 & 0.10 & 17 & Y \\ \hline
\end{tabular}}
\tablenotetext{\footnotesize{a}}{\footnotesize{The adopted rest frequency (with errors in the last digit); references: (1) \cite{Pickett1998}; (2) \cite{Muller2004}. The listed methanol transitions are indicated in the text by the frequency in GHz rounded to two decimal places.}} 
\end{table}

\begin{table}[ht]
\caption{Observation parameters: Continuum}
\label{tab:obsc}
\begin{tabular}{ccccccccc}
\hline \hline
Band\tablenotemark{\footnotesize{a}} & Freq. & Epoch & Int. Time & Synth. Beam & PA    & 1$\sigma$ rms     & \multicolumn{2}{c}{Detect.} \\ \cline{8-9} 
     & (GHz) &       & (min)     & (arcsec)    & ($^{\circ}$)   & ($\mu$Jy/beam) & MM1          & MM3\tablenotemark{\footnotesize{b}}          \\ \hline
K    & 20.0  & I     & 32        & 1.28×0.30   & +0.28 & 16         & Y            & Y            \\
     &       & II    & 19        & 0.78×0.32   & −9.98 & 15         & N            & N     \\ \hline      
\end{tabular}
\tablenotetext{\footnotesize{a}}{\footnotesize{The following maser lines were detected in the continuum window (the listed lines were flagged during data processing to avoid false detection in continuum images): 
 20.97 and 20.35 (epoch I only) GHz.}} 
 \tablenotetext{\footnotesize{b}}{\footnotesize{The source is not considered further in the present study.}}
\end{table}

\section{Results} \label{sec:results}

\subsection{Methanol Maser Emission} 

Maser emission is detected in all frequency bands: there are not only well-known methanol masers at 6.67 and 12.18~GHz, but also rare and recently discovered ones. The four newly discovered methanol maser transitions at 6.18, 12.23, 20.35, and 20.97~GHz, detected in G358.93$-$0.03 with the 26-m HartRAO telescope \citep{MacLeod19} and ATCA \citep{Breen19}, were observed at epoch II with the VLA.  Of the four, only the 20.35 GHz maser  was not detected with the VLA as it had already faded away by the date of the observation. 

\begin{figure*}[ht]
\centering
\gridline{\fig{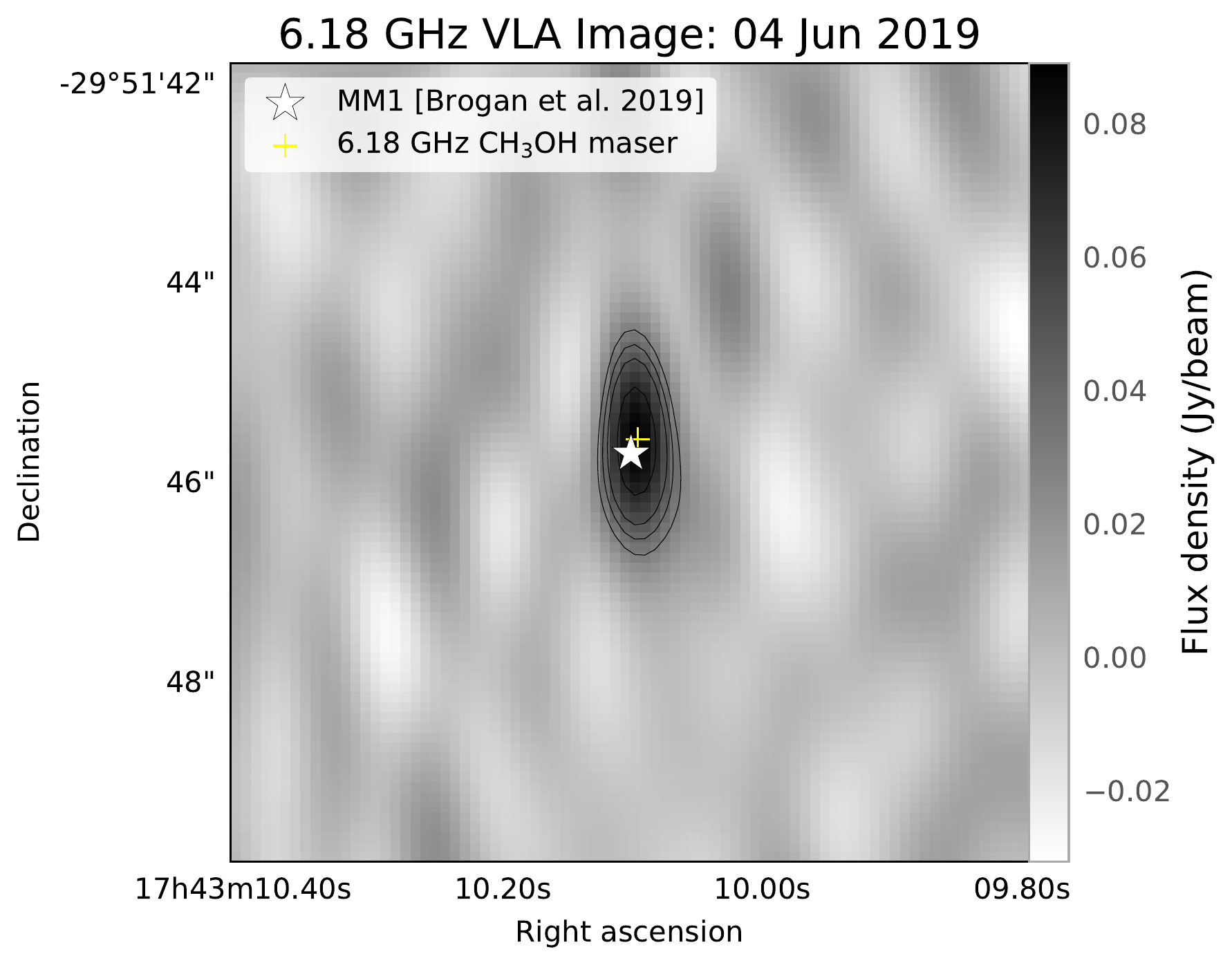}{0.5\textwidth}{(a)}
          \fig{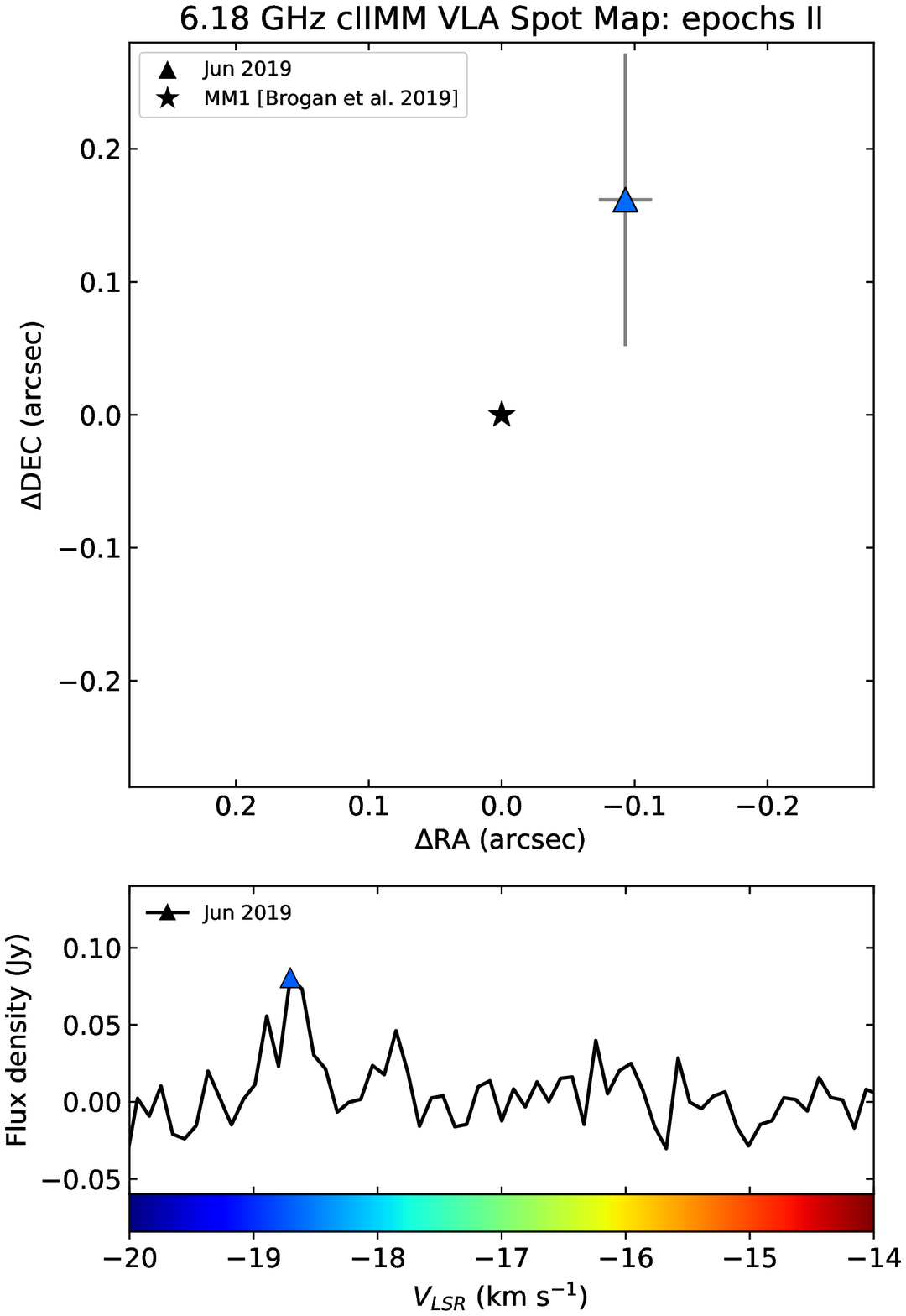}{0.3\textwidth}{(b)}
				  }
\caption{The 6.18 GHz CH$_3$OH maser emission detected in G358.93$-$0.03 with the VLA. \textit{Left panel:} (a) VLA image of the brightest 6.18 GHz CH$_3$OH maser emission spot detected at epoch II. The yellow cross marks the fitted peak of maser emission. \textit{Right panels:} (b) 6.18 GHz methanol maser spot map and spectrum (the marker on the spectrum corresponds to the maser spot on the map). Plots are color-coded by radial velocity (see colorbar for color scale). 
The gray error bar indicates the position fitting errors from Table~\ref{tab:T61GHZ}.
\label{fig:61GHZ}}
\end{figure*}

\begin{figure}[ht]
\begin{tabular}{cc}
    \begin{minipage}{0.5\textwidth} \includegraphics[width=0.8\textwidth]{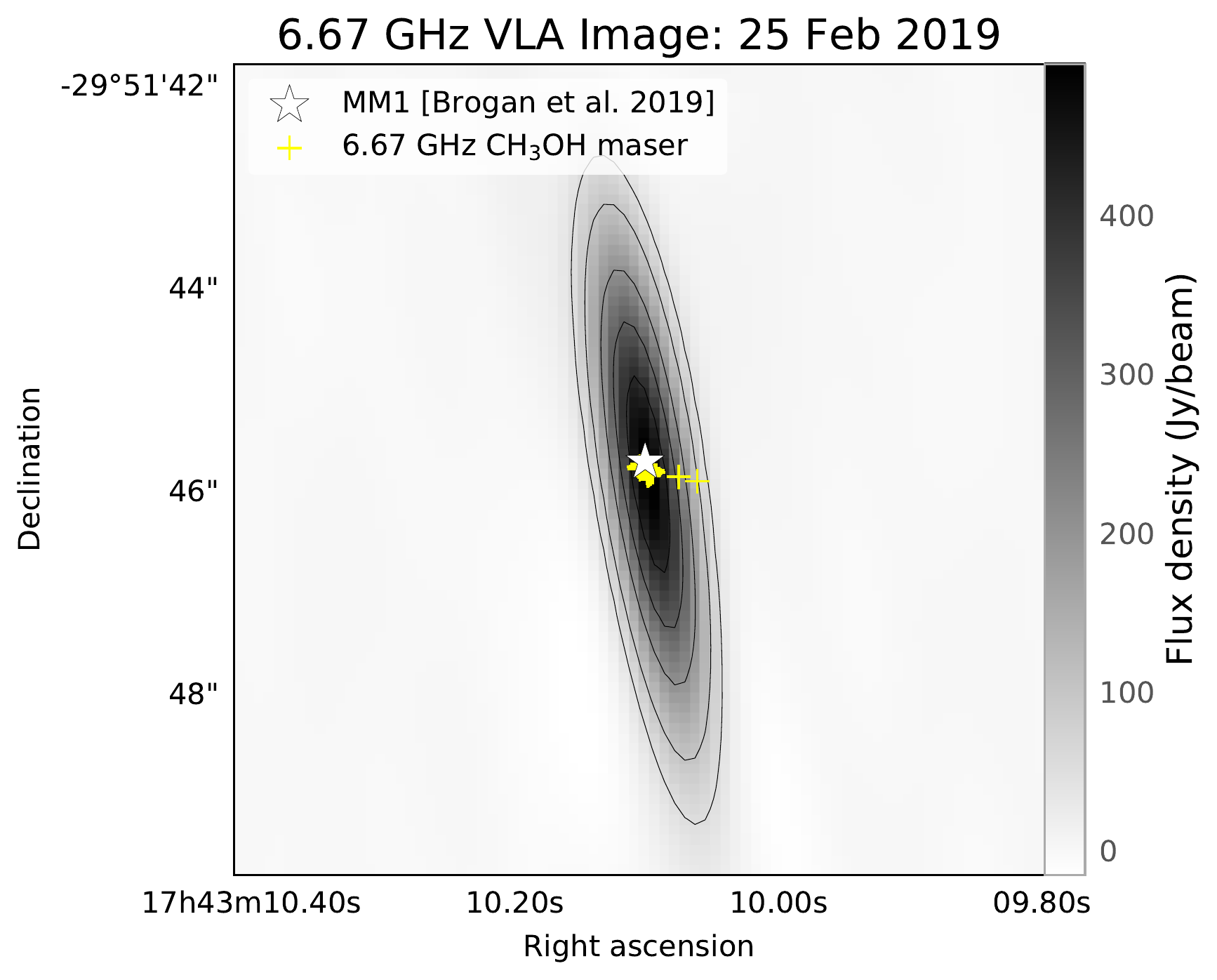} \\ 
    \includegraphics[width=0.8\textwidth]{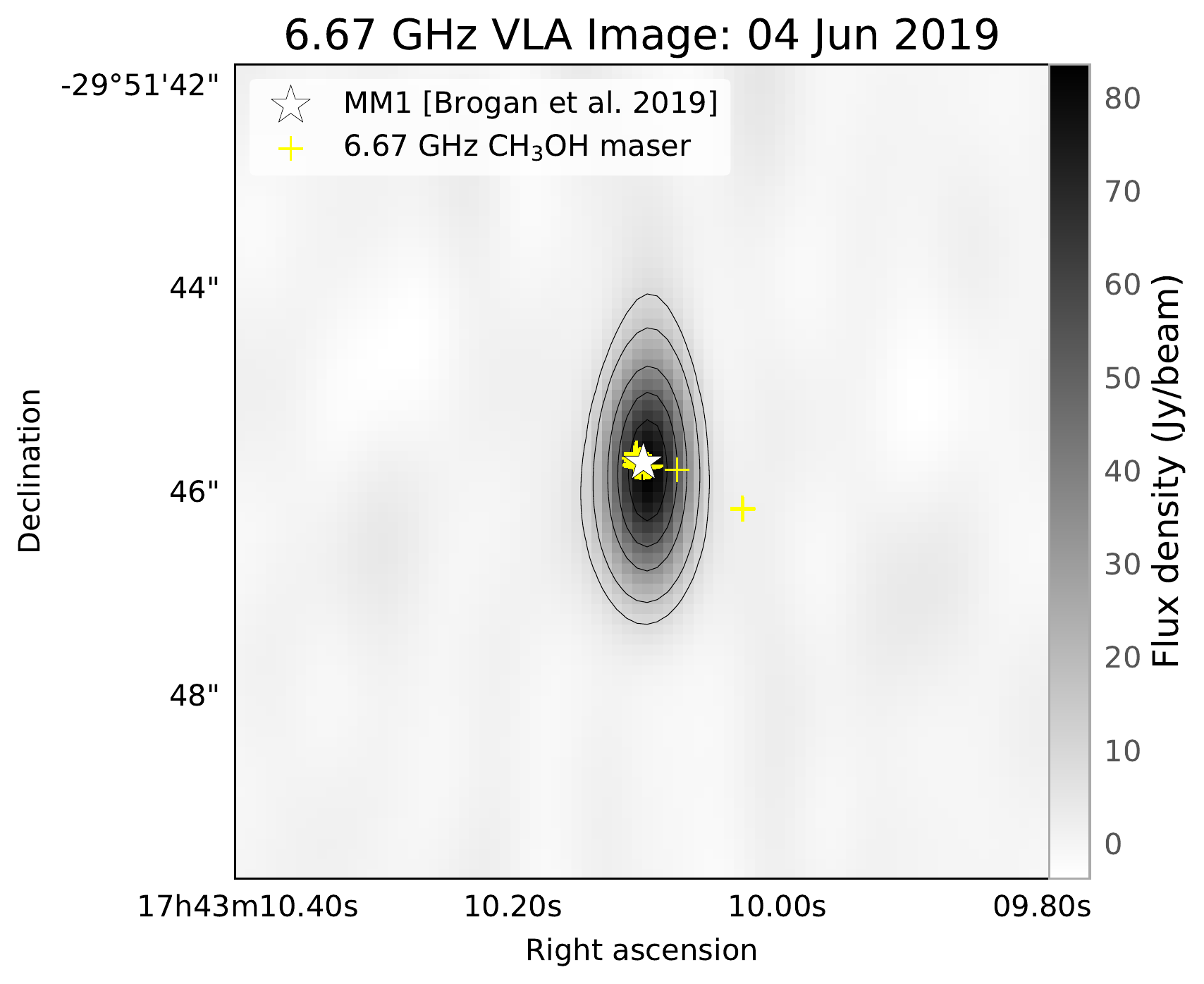} \end{minipage}
    \begin{minipage}{0.5\textwidth} \includegraphics[width=0.9\textwidth]{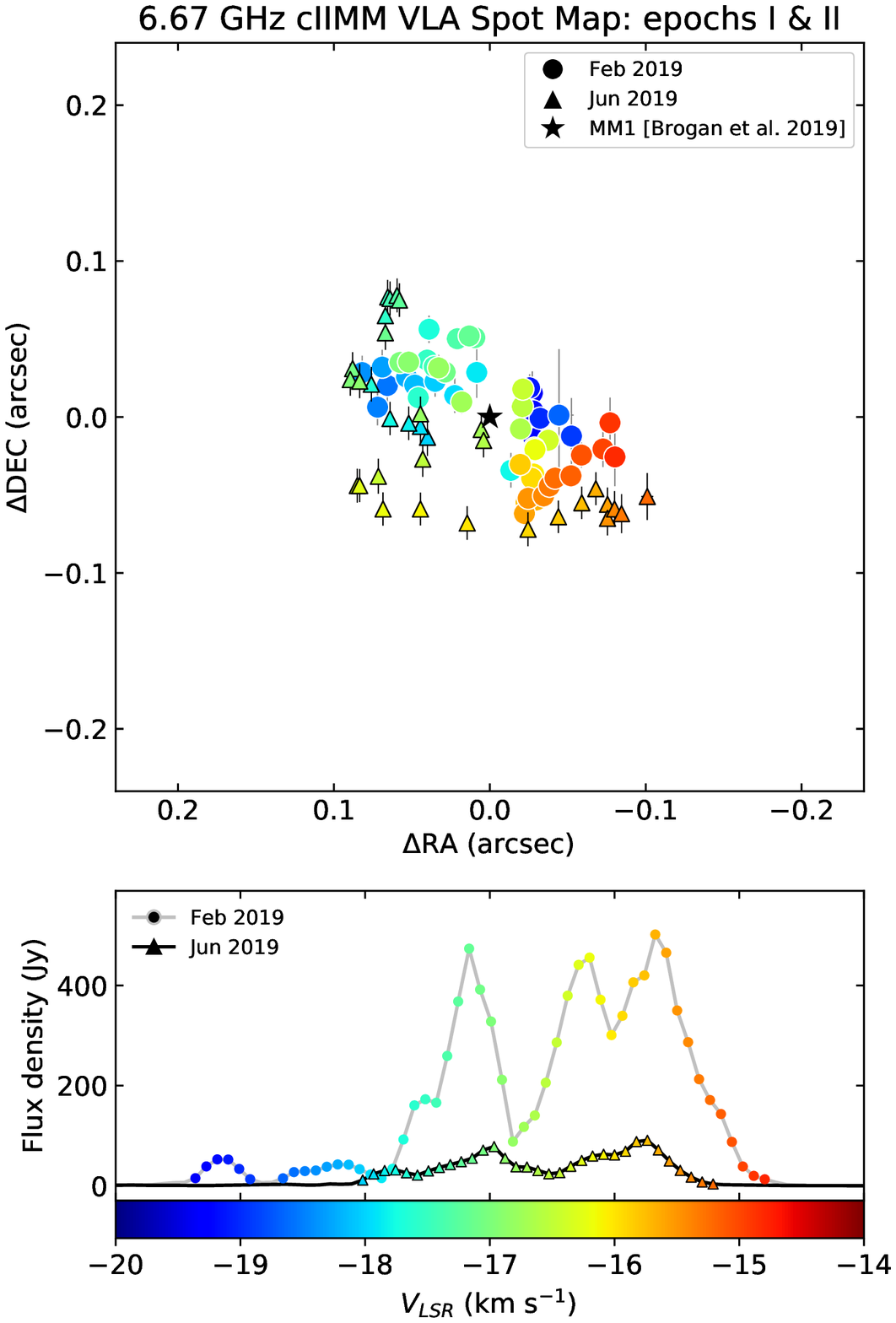} \end{minipage}& 
    \end{tabular}
        \caption{The 6.67 GHz CH$_3$OH maser emission detected in G358.93$-$0.03 with the VLA. \textit{Left panels:} VLA images of the brightest 6.67 GHz CH$_3$OH maser emission spots detected at (a) epoch I and (b) epoch II. Yellow crosses mark the peaks of maser emission (maser spots). \textit{Right panels:} (c) superimposed 6.67 GHz methanol maser spot maps and spectra (the markers on the spectra correspond to the maser spots on the map) from the epochs I and II. Plots are color-coded by radial velocity (see colorbar for color scale). The error bars indicate the position fitting errors from Table~\ref{tab:T67GHZ}. \label{fig:67GHZ}}
    \end{figure}

\begin{figure}[ht]
\begin{tabular}{cc}
    \begin{minipage}{0.5\textwidth} \includegraphics[width=0.8\textwidth]{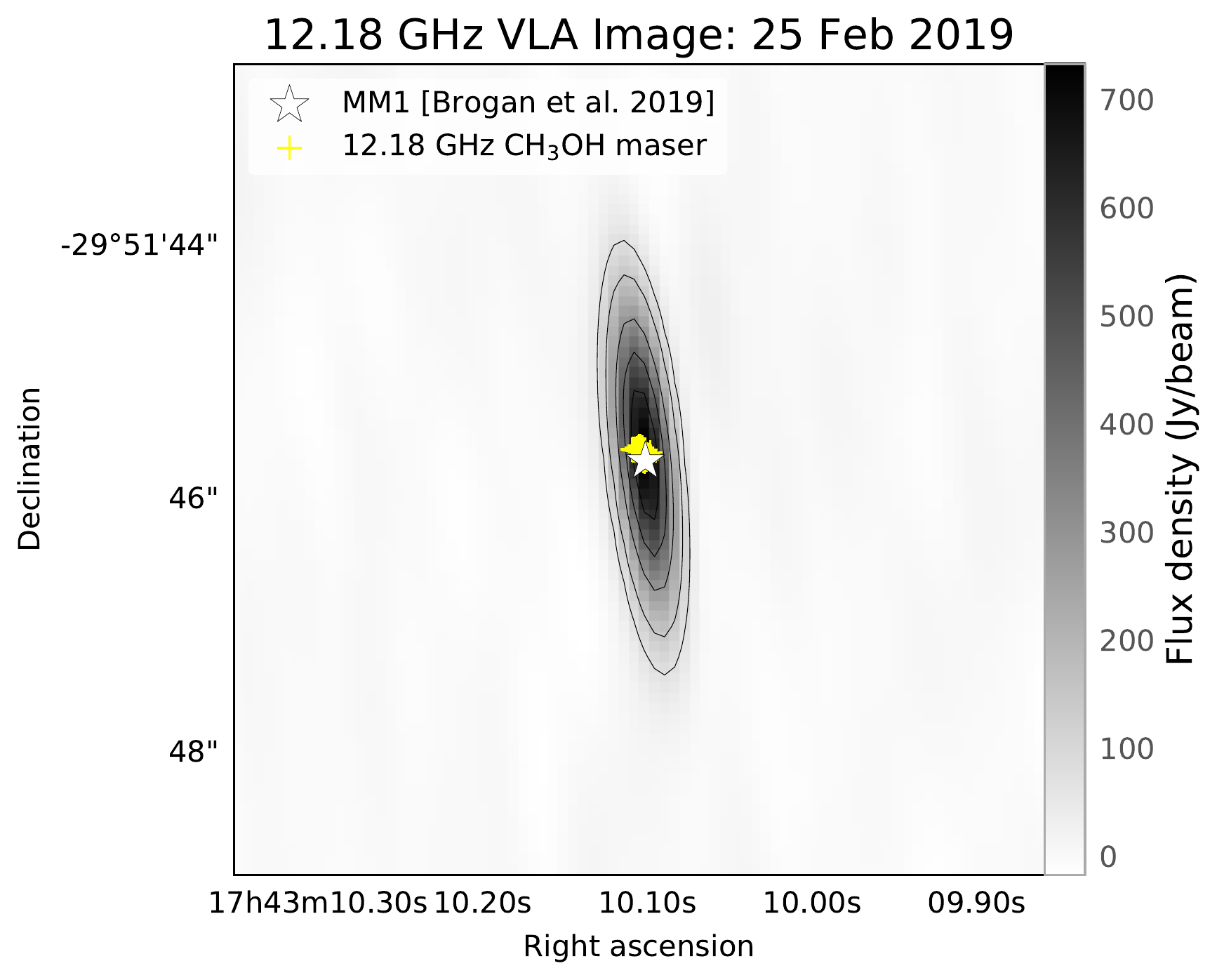} \\ 
    \includegraphics[width=0.8\textwidth]{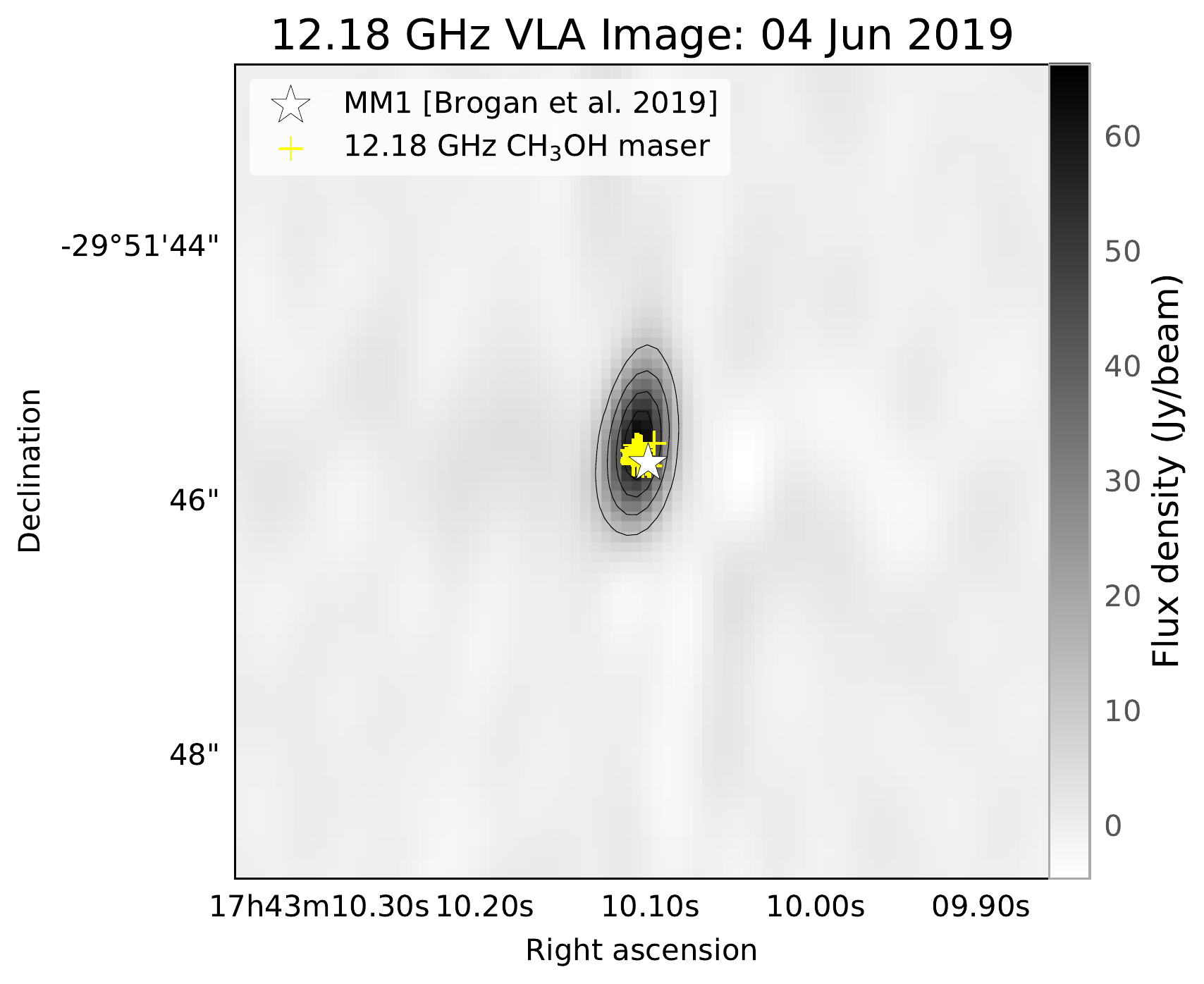} \end{minipage}
    \begin{minipage}{0.5\textwidth} \includegraphics[width=0.9\textwidth]{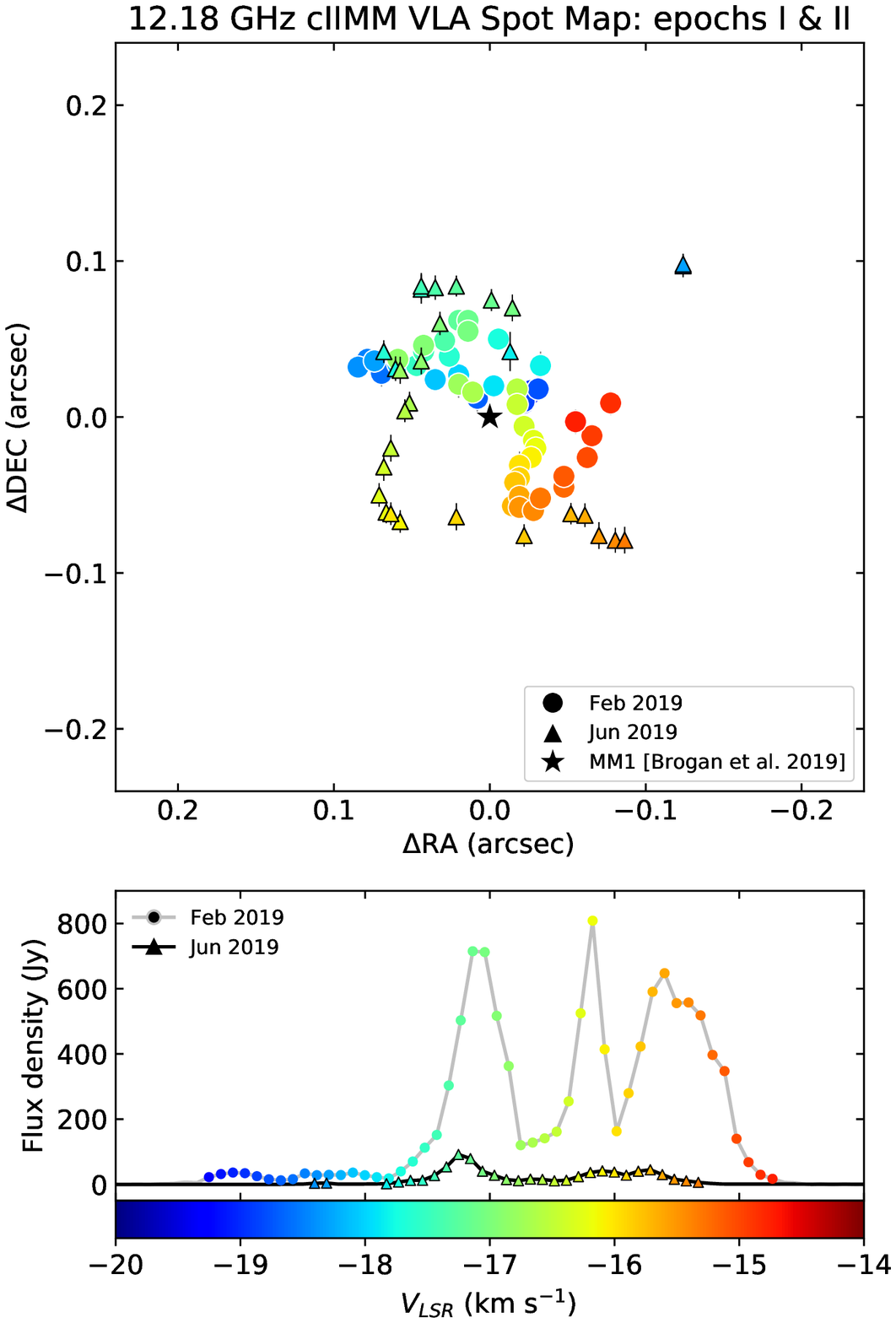} \end{minipage}& 
    \end{tabular}
        \caption{Same as Figure~\ref{fig:67GHZ} but for the 12.18 GHz CH$_3$OH maser.
\label{fig:12GHZ}}
\end{figure}

\begin{figure*}[ht]
\gridline{\fig{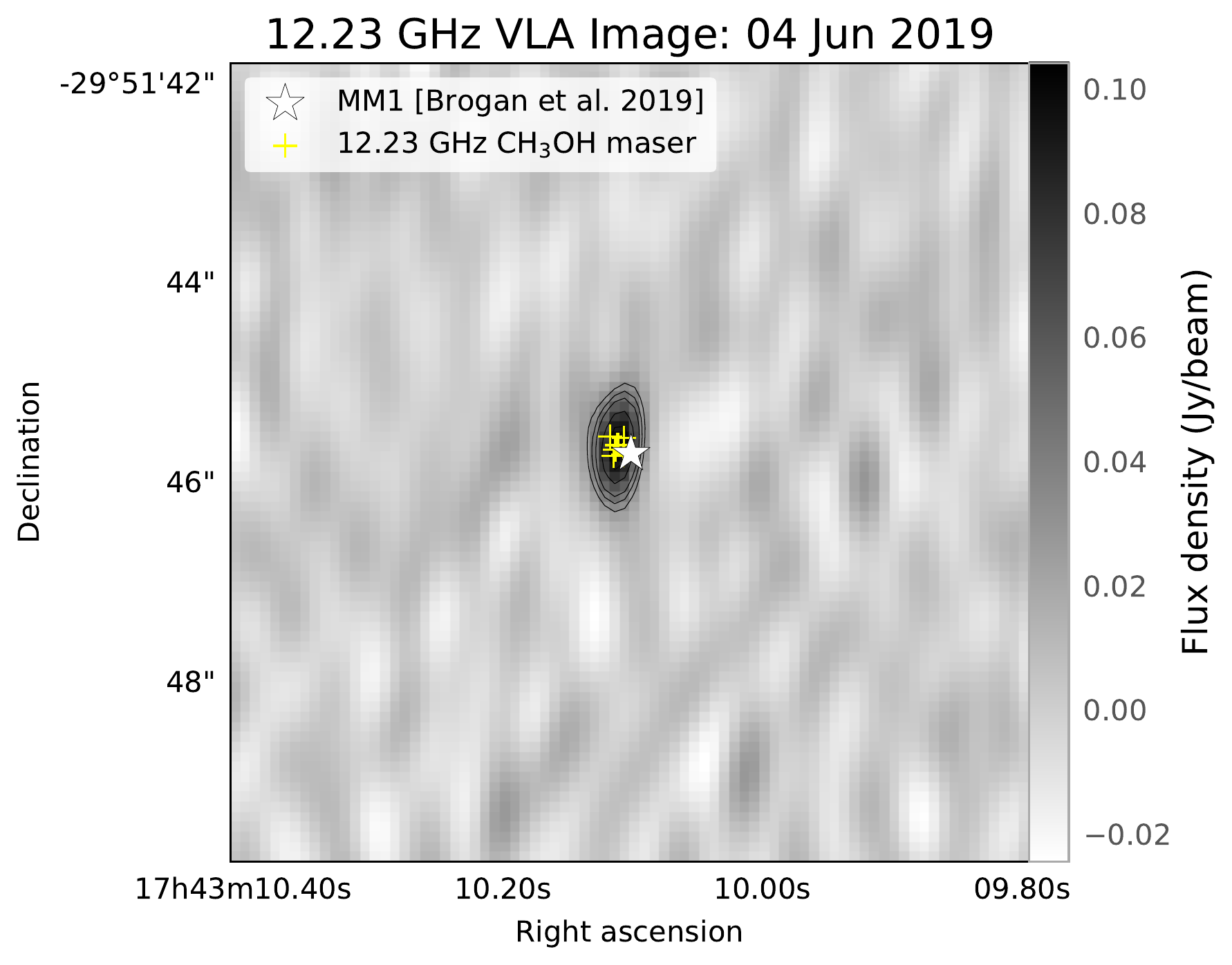}{0.5\textwidth}{(a)}
          \fig{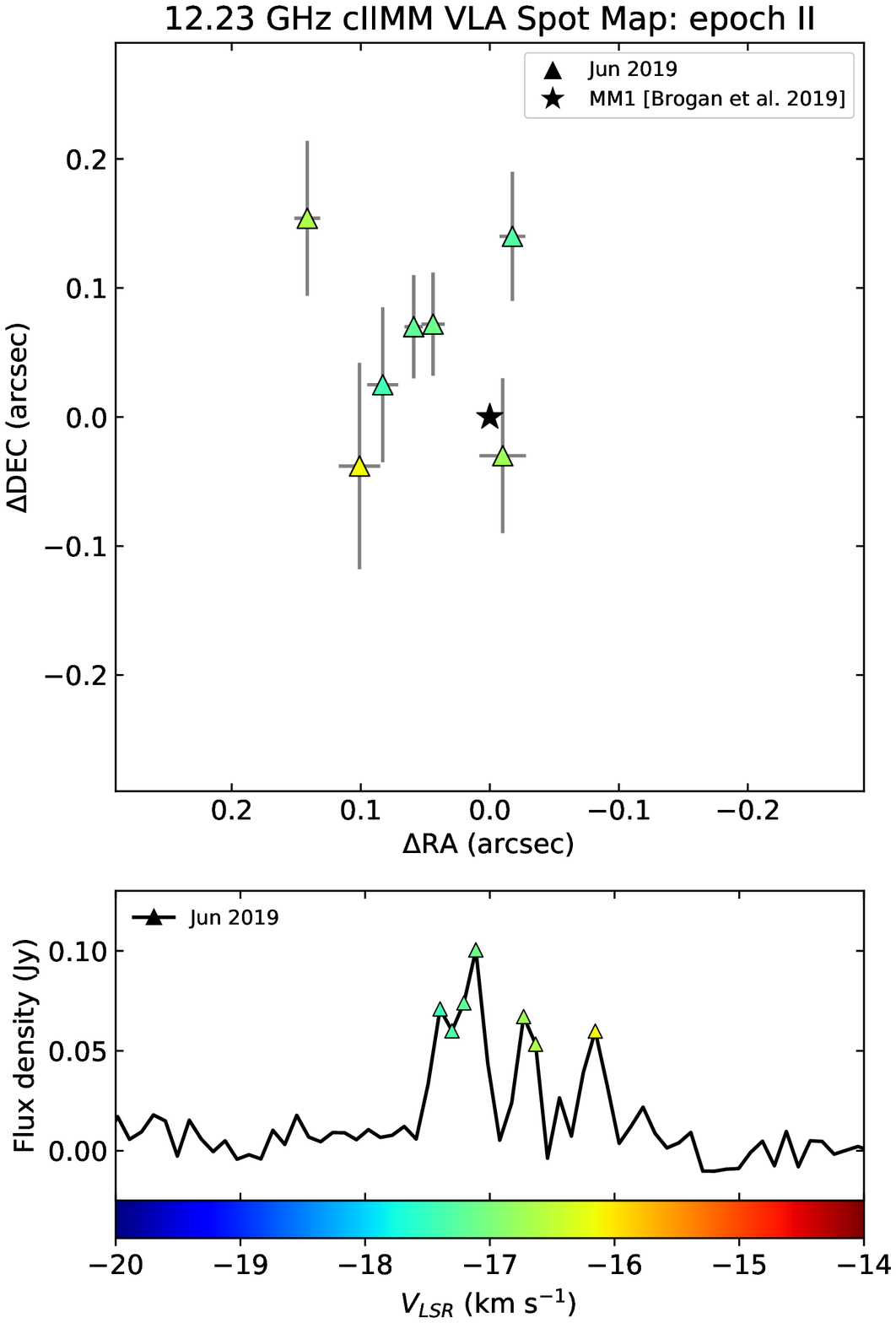}{0.28\textwidth}{(b)}
				  }
\caption{Same as Figure~\ref{fig:61GHZ} but for the 12.23 GHz CH$_3$OH maser.
\label{fig:122GHZ}}
\end{figure*}

\begin{figure*}[ht]
\gridline{\fig{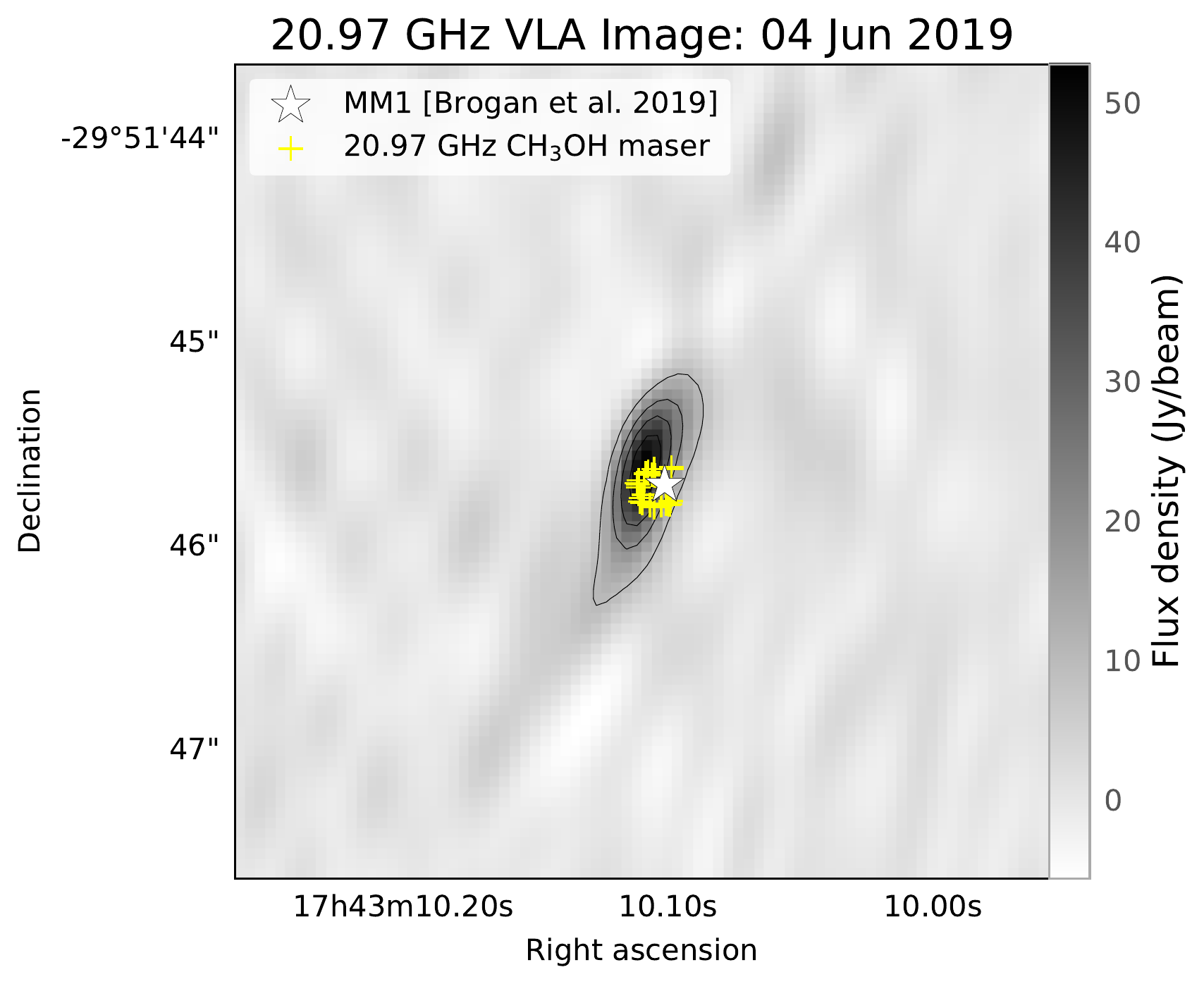}{0.5\textwidth}{(a)}
          \fig{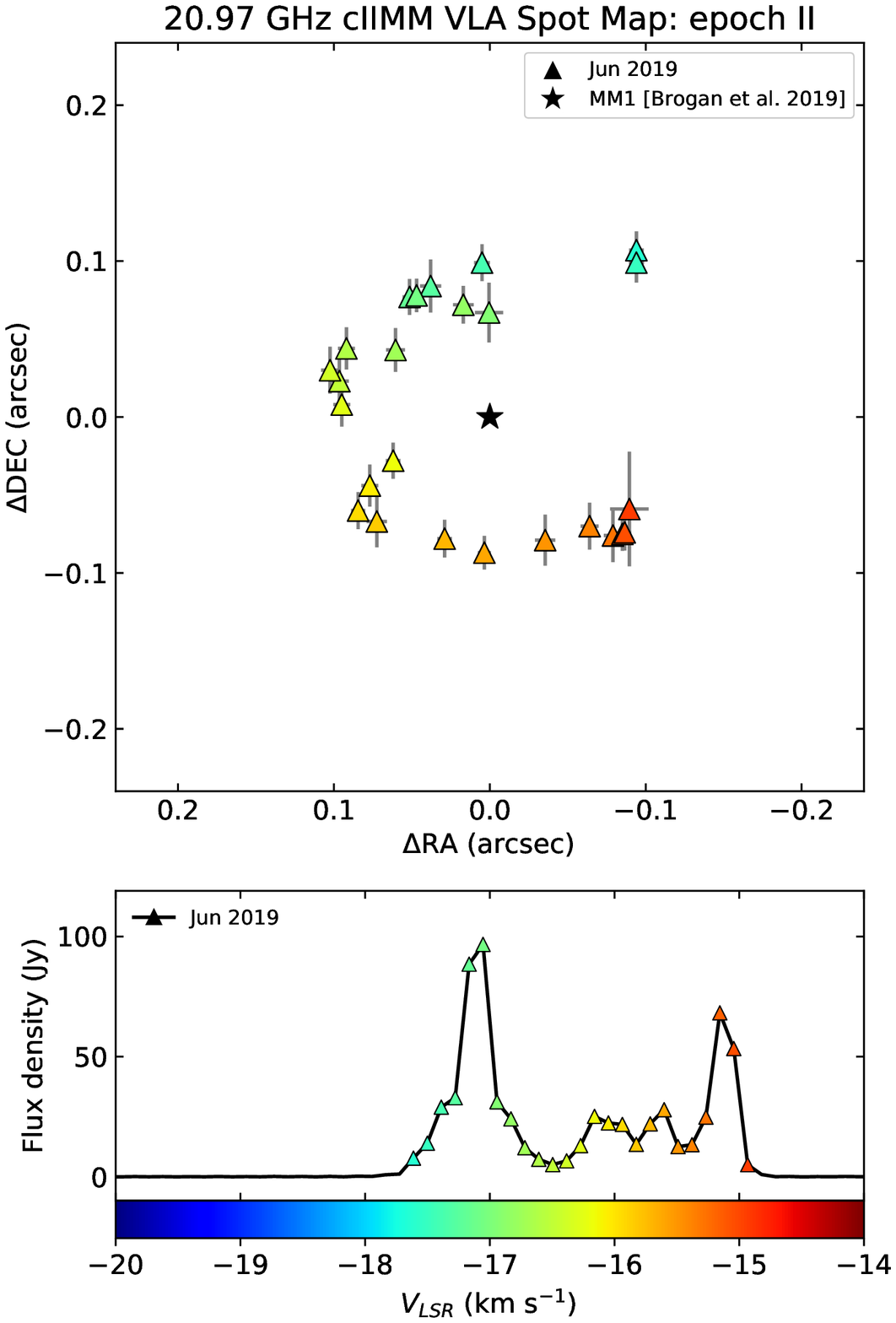}{0.28\textwidth}{(b)}
				  }
\caption{Same as Figure~\ref{fig:61GHZ} but for the 20.97 GHz CH$_3$OH maser.
\label{fig:20GHZ}}
\end{figure*}

\begin{figure}[ht]
\begin{tabular}{cc}
    \begin{minipage}{0.5\textwidth} \includegraphics[width=0.8\textwidth]{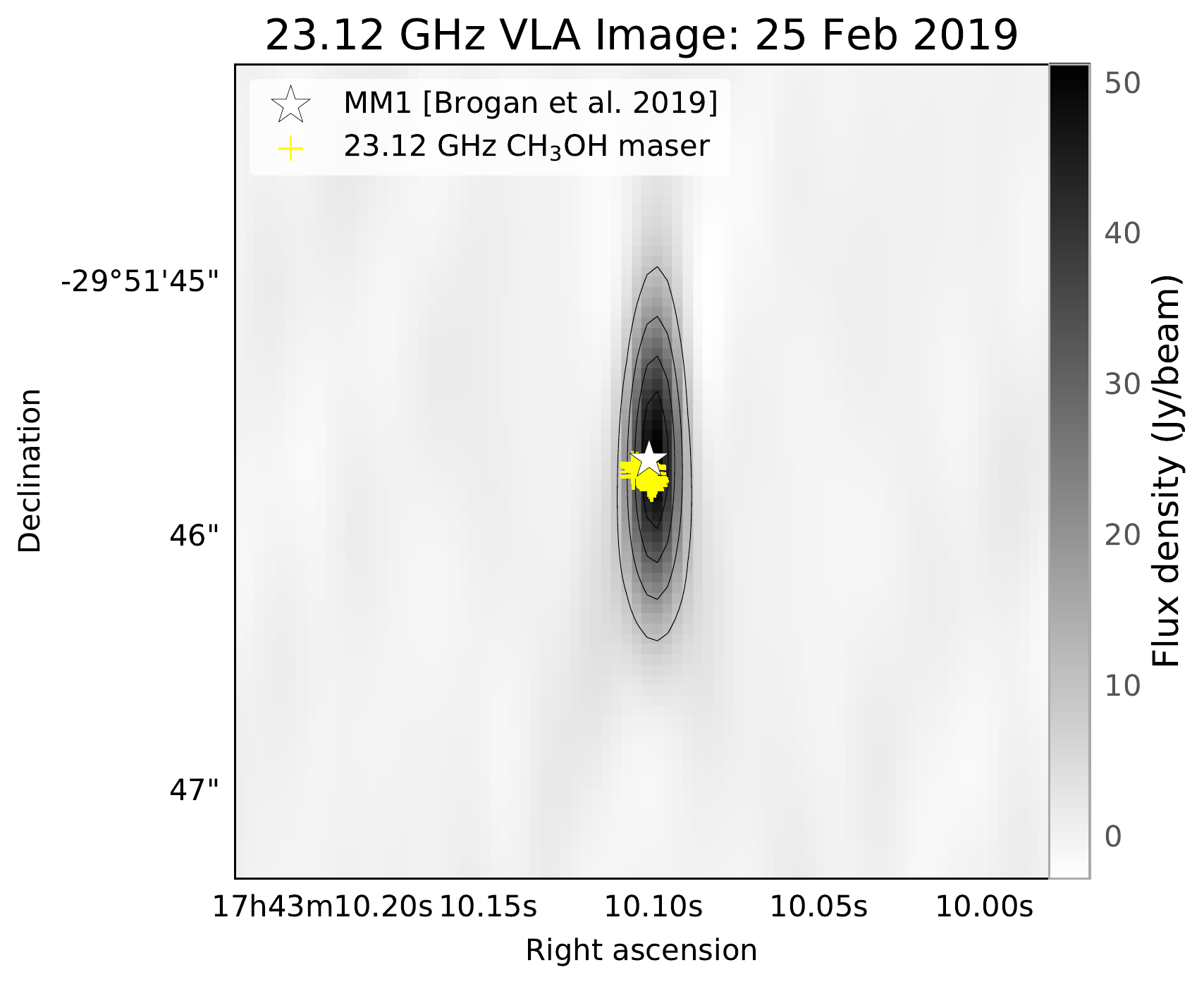} \\ 
    \includegraphics[width=0.8\textwidth]{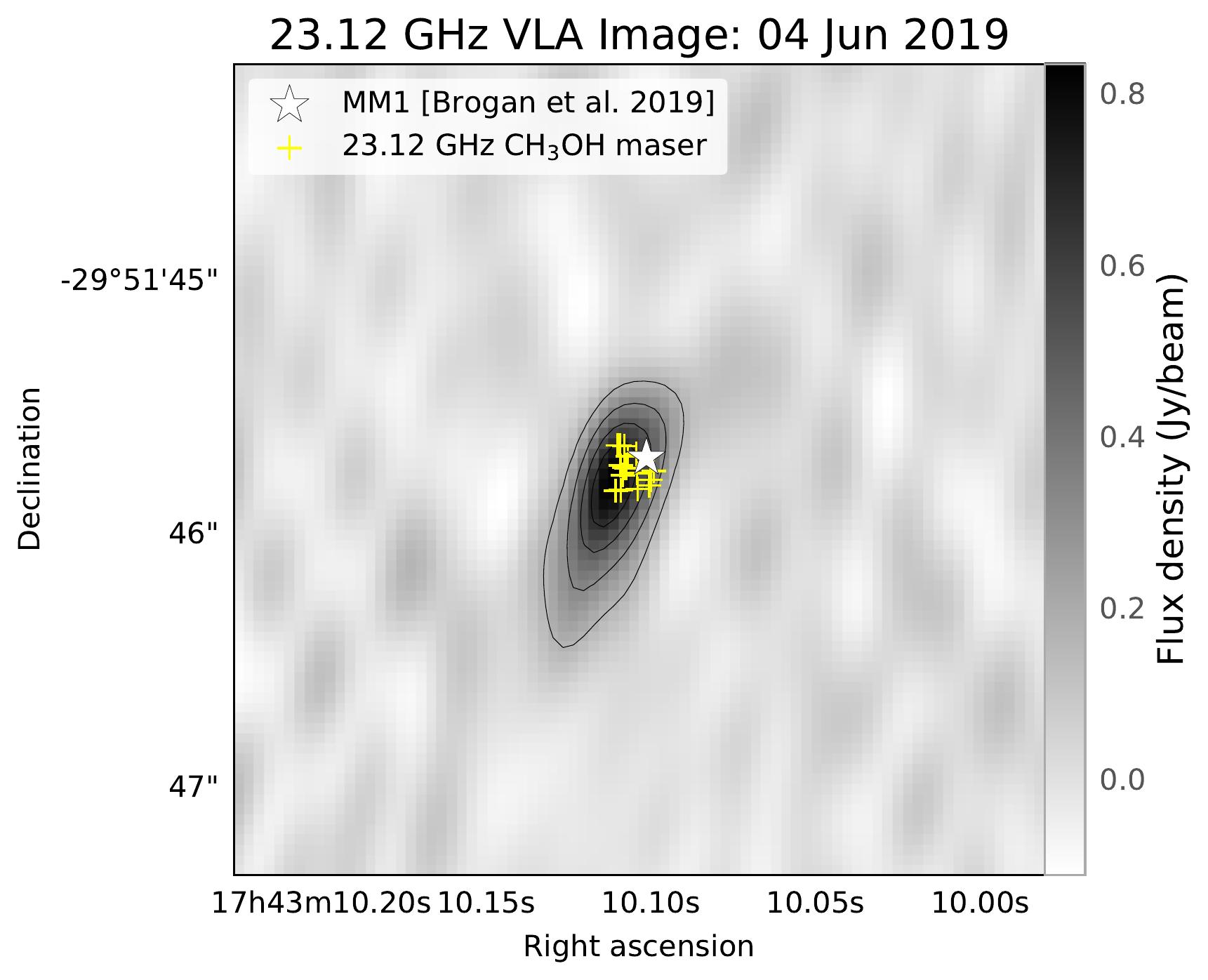} \end{minipage}
    \begin{minipage}{0.5\textwidth} \includegraphics[width=0.9\textwidth]{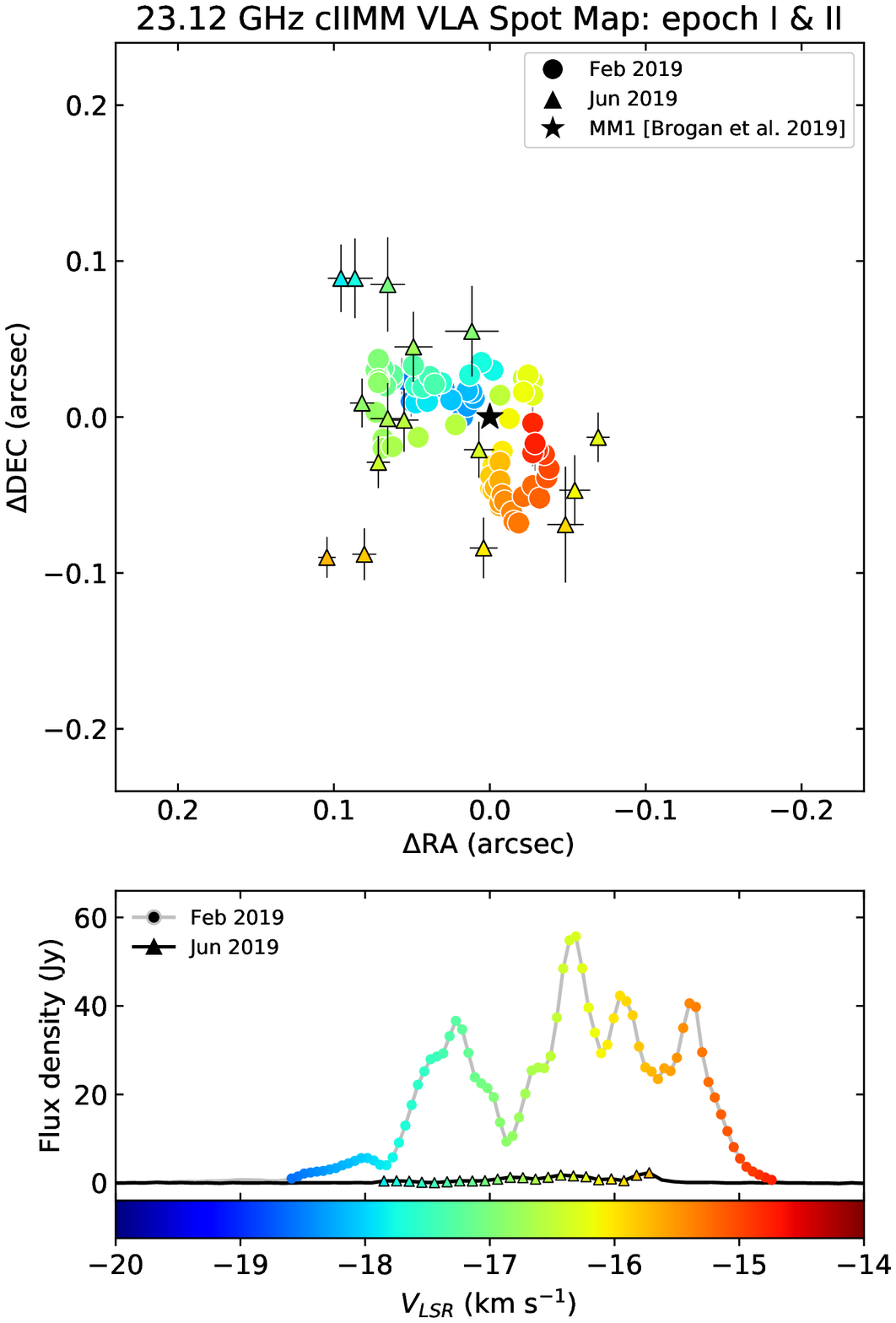} \end{minipage}& 
    \end{tabular}
        \caption{Same as Figure~\ref{fig:67GHZ} but for the 23.12 GHz CH$_3$OH maser.
\label{fig:23GHZ}}
\end{figure}

The position, velocity, integrated and peak flux density of each maser spot are listed in Tables \ref{tab:T61GHZ}--\ref{tab:T23GHZ}. These tables, in machine-readable format, are presented in the Appendix. Images of the brightest emission spots detected at a particular frequency, as well as spectra of these maser transitions 
and their spatial distribution, are presented in Figures~\ref{fig:61GHZ}--\ref{fig:23GHZ}. Each map is centered on the position of the millimeter core MM1 from \cite{Brogan19} (indicated by a star marker).

Note that the structures seen in the maser spot maps are much smaller than the angular size of the VLA synthesized beam in each band (Table \ref{tab:obsspl}). 
The high signal-to-noise achieved for the bright maser emission allows us to fit the maser spot positions with sub-beamsize accuracy. Nevertheless, if there is more than one maser spot in a velocity channel, the spatial and velocity structure will be dominated by the brighter components (i.e., we obtain 'centroid mapping'; see Section \ref{sec:other} for the further discussion on the issue). This is especially true for the crowded maser regions detected at the flare epoch. \\

\newpage
\textbf{6.18~GHz.} The 6.18 GHz class II methanol maser is one of the  torsionally excited lines discovered with the ATCA during the burst in G358.93$-$0.03 \citep{Breen19}. By the time of the VLA epoch II observations, the flux density of the 6.18 GHz maser had dropped significantly and remained at the $\sim$0.1~Jy level (Figure~\ref{fig:61GHZ}b), compared to the flaring flux density of $\sim$300~Jy. At the VLA epoch II, only the $-$18.6 km s$^{-1}$ feature remained detectable. The ATCA spectrum showed emission from $-18.8$ to $-15.5$~km~s$^{-1}$ with a peak emission at $-$16.2 km s$^{-1}$. However, both the ATCA and VLA positions of the 6.18 GHz maser coincide within their positional uncertainties.

\textbf{6.67 GHz.} The first epoch of the VLA observations was conducted soon after the first 6.67 GHz maser flare in G358.93$-$0.03. 
On  February 25, 2019 (VLA epoch I), the single-dish flux density of the velocity feature at $\sim-16$~km~s$^{-1}$  had decreased from $\sim$600~Jy to $\sim$450~Jy, according to iMet data\footnote{\url{http://vlbi.sci.ibaraki.ac.jp/iMet/G358.9-00-190114/daily.html}}. We note that February 25 was also the date when the dominant feature at V$_{LSR}$ $\sim-16$~km~s$^{-1}$ was overtaken by the feature at V$_{LSR}$ $\sim-17$~km~s$^{-1}$.

The brightest flare of the 6.67 GHz maser occurred on March 12, when the $\sim-17$~km~s$^{-1}$  
velocity feature reached 900~Jy.
By the second VLA epoch (June 4, 2019), the single-dish flux density of the 6.67 GHz methanol maser had decreased to $\sim$90~Jy, with both lines --- at V$_{LSR}$ $\sim-$16 km/s (dominant in the first flare) and $\sim-$17 km/s (dominant in the second flare) --- having about the same amplitude (see Fig.~\ref{fig:timeline} and \ref{fig:67GHZ}).

The methanol maser emission at 6.67 GHz is detected at both VLA epochs, coming from an area of $\sim0.2^{\prime\prime}$ ($\sim$1000~AU, for an assumed distance of 6.75 kpc) around the MM1 position (Fig. \ref{fig:67GHZ}). Several fainter ($<$1~Jy) maser spots were detected outside of the main cluster. These ``distant'' maser spots are located to the west of the main cluster (Fig. \ref{fig:67GHZ}(a,b)).
At the first VLA epoch, the cluster of the 6.67 GHz maser spots was elongated in the NE-SW direction, with most of the blue-shifted masers to the NE and the red-shifted features to the SW.  Here, 'blue' and 'red' refer to the centroid of the velocity range, of about $-17$ km~s$^{-1}$.
In contrast, the 6.67 GHz maser spots detected at the second epoch are arranged in a bow-shaped structure, again with the blue-shifted velocity features to the north and red-shifted features to the south.
The overall velocity pattern of the spectrum did not change greatly between epochs, although the weaker velocity features at V$_{LSR}$ $\sim-18$ -- $-19.5$ km s$^{-1}$ detected in the first epoch were not detected at the second epoch.


\textbf{12.18 GHz.} The methanol maser emission at 12.18 GHz is found in a region of $\sim$0.2$^{\prime\prime}$ around MM1 (Figure \ref{fig:12GHZ}). This maser transition shows a  behaviour quite similar to that of the 6.67 GHz maser.
At the VLA epoch I, the 12.18 GHz maser emission is also elongated in the NE-SW direction with blue-shifted velocity features to the north-east and red-shifted features to the south-west. At the VLA epoch II, there is a wide bow-shaped structure with red-shifted masers to the north and a less-ordered structure to the south, comprised of blue-shifted masers.

At the post-flare epoch (VLA epoch II), both maser transitions faded to about the same flux density with the 6.67 GHz maser slightly stronger: the peak flux densities of the 6.67 GHz and 12.18 GHz emission were $\sim$84 Jy and $\sim$79 Jy, respectively.


\newpage 
\textbf{12.23~GHz.} The 12.23~GHz methanol maser was discovered in observations of G358.93$-$0.03 with the 26-m HartRAO telescope \citep{MacLeod19}. The emission in this transition was detected in March 2019 with a flux density of $\sim$1100~Jy, but only $\sim$0.1~Jy remained at the VLA observation in June 2019 (Fig. \ref{fig:122GHZ}). 
The flaring spectral feature at V$_{LSR}$ $-$15.5 km s$^{-1}$ detected at HartRAO was below our 3$\sigma$ detection limit, and, as for the 6.18 GHz maser, the only emission still detectable at epoch II has a velocity close to the minimum velocity of the flaring spectrum. The brightest 12.23~GHz maser spot detected with the VLA is at V$_{LSR}$ $-$17.1 km s$^{-1}$. Maser spots detected with the VLA lie in a region of $\sim$0.2$^{\prime\prime}$ slightly to the NE of MM1.

\newpage
\textbf{20.97 GHz.} The 20.97 GHz methanol maser was discovered by the Mopra observations of G358.93$-$0.03 conducted during the burst \citep{Breen19}. In the present study, this methanol maser transition was observed only at the second epoch. The maser emission at 20.97 GHz occupies the velocity range of $\sim-18$ to $-15$ km~s$^{-1}$, the same as other methanol maser transitions detected in the source.  By the time of the VLA observations, the flaring 20.97 GHz spectral feature at V$_{LSR}$ $\sim-15$~km~s$^{-1}$ had decayed from $\sim$1000~Jy \citep{Breen19} to $\sim$70~Jy (Fig. \ref{fig:20GHZ}). The blue-shifted feature at V$_{LSR}$ $\sim-17$~km~s$^{-1}$  with a flux density of $\sim$100~Jy had become the dominant one. Spatially, the maser spots trace an almost round structure of size of $\sim$0.2$^{\prime\prime}$, pointing eastward from the MM1 position. There is a clear velocity gradient with blue-shifted features to the north and red-shifted features to the south.

\newpage
\textbf{23.12 GHz.} During the burst, the 23.12 GHz methanol masers associated with G358.93$-$0.03 showed the brightest and the most complex spectrum  \citep{MacLeod19} of this rare maser transition 
detected to date (see \citealt{Galvan10} and references therein). The 23.12 GHz emission covers the same velocity range as of the 6.67 and 12.18 GHz masers, but with about $10\times$ lower flux density. By the VLA epoch II, the 23.12 GHz maser had faded to $\sim$2 Jy, but continued to have a complex, multi-component spectrum.

Spatially, the 23.12 GHz maser emission is found around
MM1 (Figure \ref{fig:23GHZ}). 
At the flare epoch, the 23.12 GHz maser cluster appeared to be more compact than the 6.67/12.18 GHz masers, with a size of $\sim$0.15$^{\prime\prime}$. 
The flaring 23.12 GHz emission consists of two elongated sub-clusters, with the northern cluster hosting blue-shifted velocity features and the southern cluster with red-shifted features. At VLA epoch II, the 23.12 GHz emission is found in an expanded region of $\sim$0.2$^{\prime\prime}$ (note that the lower flux densities at epoch II result in higher positional uncertainties, which could affect the apparent size of the region). The structure of the masing region at the post-flare epoch is less ordered, but shows the same north-south velocity gradient.


\newpage
\subsection{Continuum Emission}

The VLA observations toward G358.93$-$0.03 detected two continuum sources (Table \ref{tab:obsc}).
The hot core MM3 showed continuum emission in all three frequency bands at both epochs. Nevertheless, MM3 has no evident association with the bursting source and the detected methanol masers, and is not considered further in the present study. The bursting source MM1 was detected in K-band at the first epoch of the VLA observations. No K-band emission above the 3$\sigma$ noise level was found at the second epoch. 

In our VLA observation, conducted on February 25, 2019, during the burst epoch, MM1 showed an integrated flux density of $\sim$0.2~mJy at 20 GHz.

Note that the only centimetre observations found in the literature on the pre-burst epoch are those of \cite{Hu16}, who report a null detection at C-band, but based on only 20 seconds of data. The latest VLA B-array observation of \cite{Chen20a} also reported a non-detection of Ku-band continuum emission in the direction of MM1 at 3$\sigma$ level of 20 $\mu$Jy.

At the post-burst epoch, on April 12, 2019 in ALMA observations, the flux density was $\sim$282~mJy at 337 GHz \citep{Brogan19}.
Note, however, that the observations were made at different stages of the source's activity, and, therefore, the results of the observations cannot be directly compared.

\begin{figure*}[ht]
\gridline{
\fig{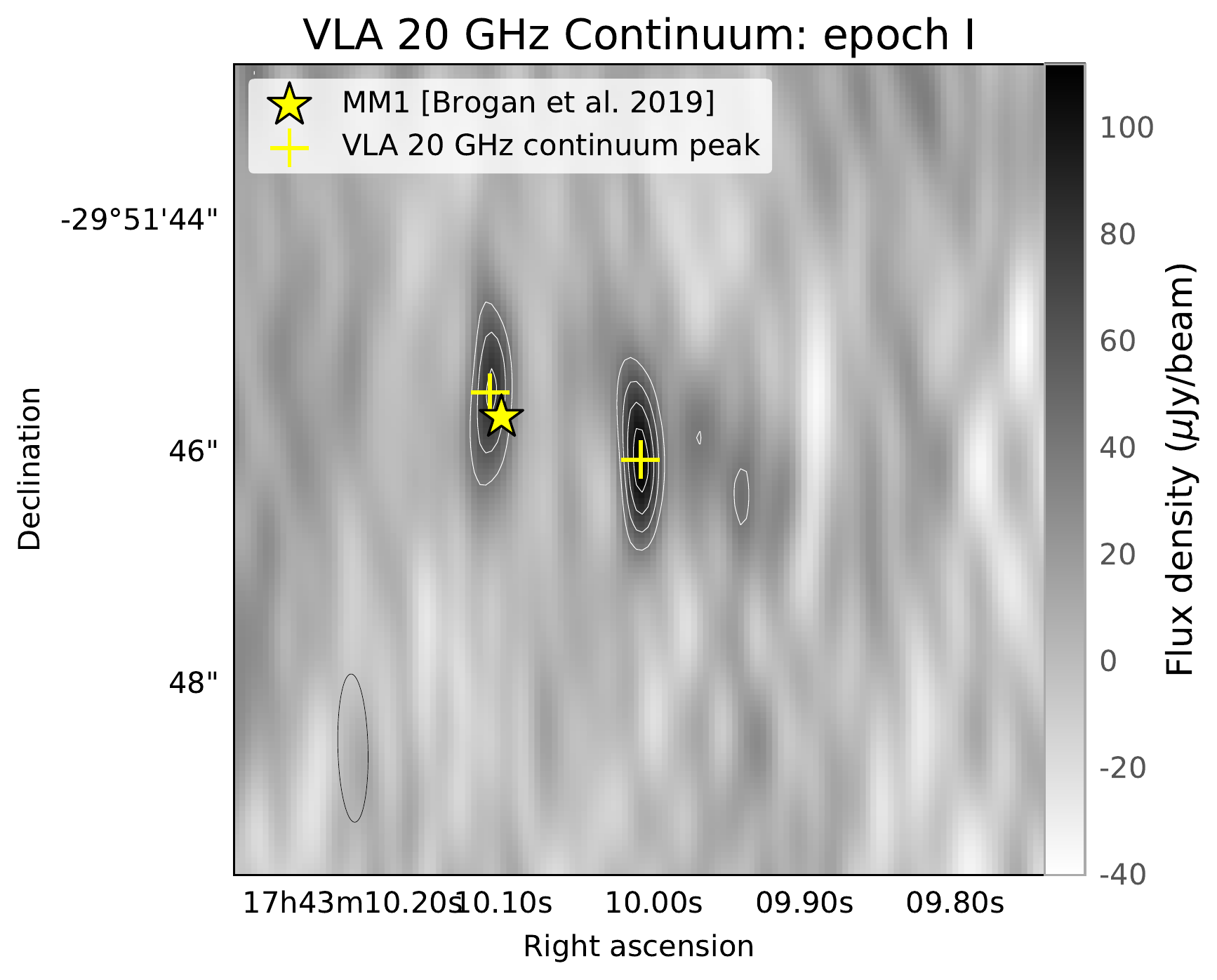}{0.45\textwidth}{}
\fig{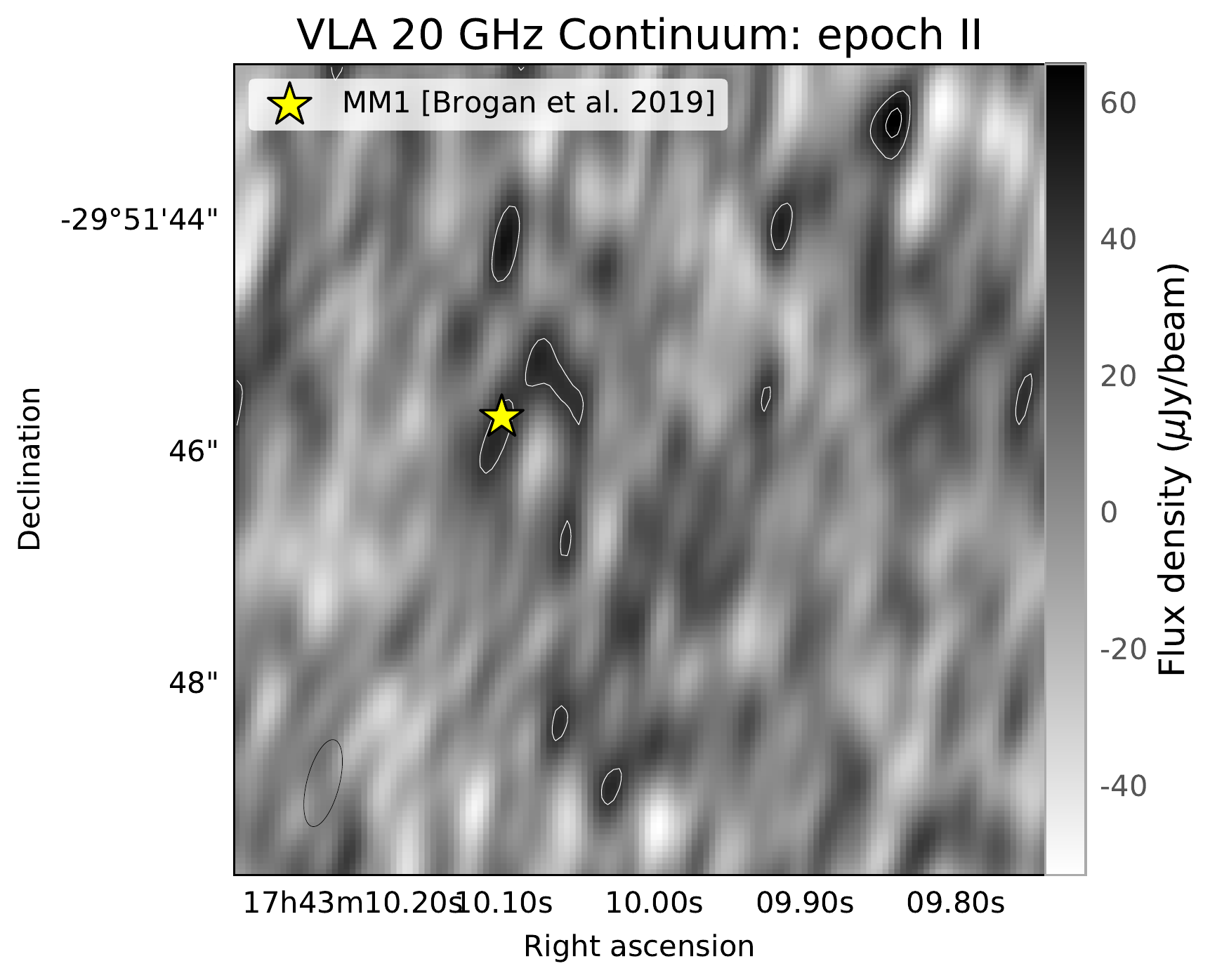}{0.45\textwidth}{}
                  }
\caption{Continuum images of G358.93$-$0.03 at 20 GHz at the first (left panels) and second (right panels) VLA epochs. Contour levels are [4, 6, 8, 10] $\times$ 10~$\mu$Jy/beam. The yellow cross indicates the detected peak(s) of continuum emission at a certain frequency. The yellow star marks the position of MM1 from \cite{Brogan19}. The synthesized beam size of an image is shown with a black ellipse in the lower left corner of each panel. \label{fig:cont}}
\end{figure*}

\begin{figure*}[ht]
\centering
\gridline{\fig{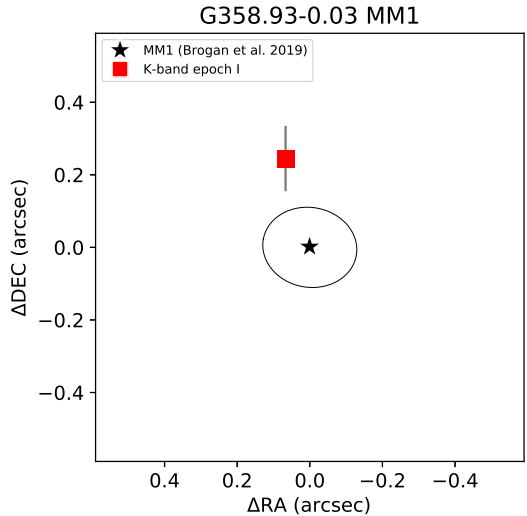}{0.4\textwidth}{}
				  }
\caption{The cm- (this work) and mm- \citep{Brogan19} continuum peak position 
for MM1. The K-band peak detected with the VLA at epoch I is shown in red. The gray error bars indicate the position fitting error 
from Table~\ref{tab:cont}. 
\label{fig:position_check}}
\end{figure*}

The K-band continuum images of MM1 at both epochs are shown in Figure \ref{fig:cont} and a summary of the parameters of the detected continuum peak is presented in Table \ref{tab:cont}. 
The  offset between the ALMA and VLA positions of MM1 is $\sim$0.2$\arcsec$ (Fig. \ref{fig:position_check}) which is similar to  the absolute positional uncertainty reported in \cite{Brogan19}.

\startlongtable
\begin{deluxetable*}{cccccccc}
\tablewidth{400pt}
\tablecaption{VLA continuum emission peak parameters \label{tab:cont}}
\tablehead{
\colhead{Association} &
\colhead{Band} & 
\colhead{Epoch} &
\colhead{RA(J2000)} & \colhead{DEC(J2000)} & \colhead{Integrated flux} & \colhead{Peak flux} & \colhead{SNR} \\
\colhead{} &
\colhead{} &  \colhead{} &
\colhead{($^h$~$^m$~$^s$)} & \colhead{($^\circ$~$\arcmin$~$\arcsec$)} & 
\colhead{($\mu$Jy)} & \colhead{($\mu$Jy/beam)} & \colhead{} 
}
\startdata
MM1 & K & I &  17:43:10.106(0.0001)\tablenotemark{\footnotesize{a}}  &  $-$29:51:45.45(0.03)\tablenotemark{\footnotesize{a}} & 246(18) & 223(7) & 14\\ \hline
\enddata
\tablenotetext{\footnotesize{a}}{\centering\footnotesize{The positional uncertainties are statistical errors of fit.}} 
\end{deluxetable*}

\newpage
\section{Discussion} \label{sec:discussion}

\subsection{The accretion burst}

A unique feature of the burst in G358.93$-$0.03 is the discovery of new maser transitions \citep{Breen19,Brogan19,MacLeod19}. 
Three of these, the maser transitions at 6.18, 12.23, and 20.35 GHz, were not even predicted by maser pumping calculations performed up to now.
Others had been predicted in theoretical works (e.g., \citealt{Sobolev97b,Cragg05}), but had not been detected previously. The reasons for their rarity are not entirely clear. 

\begin{figure*}[ht]
\gridline{\fig{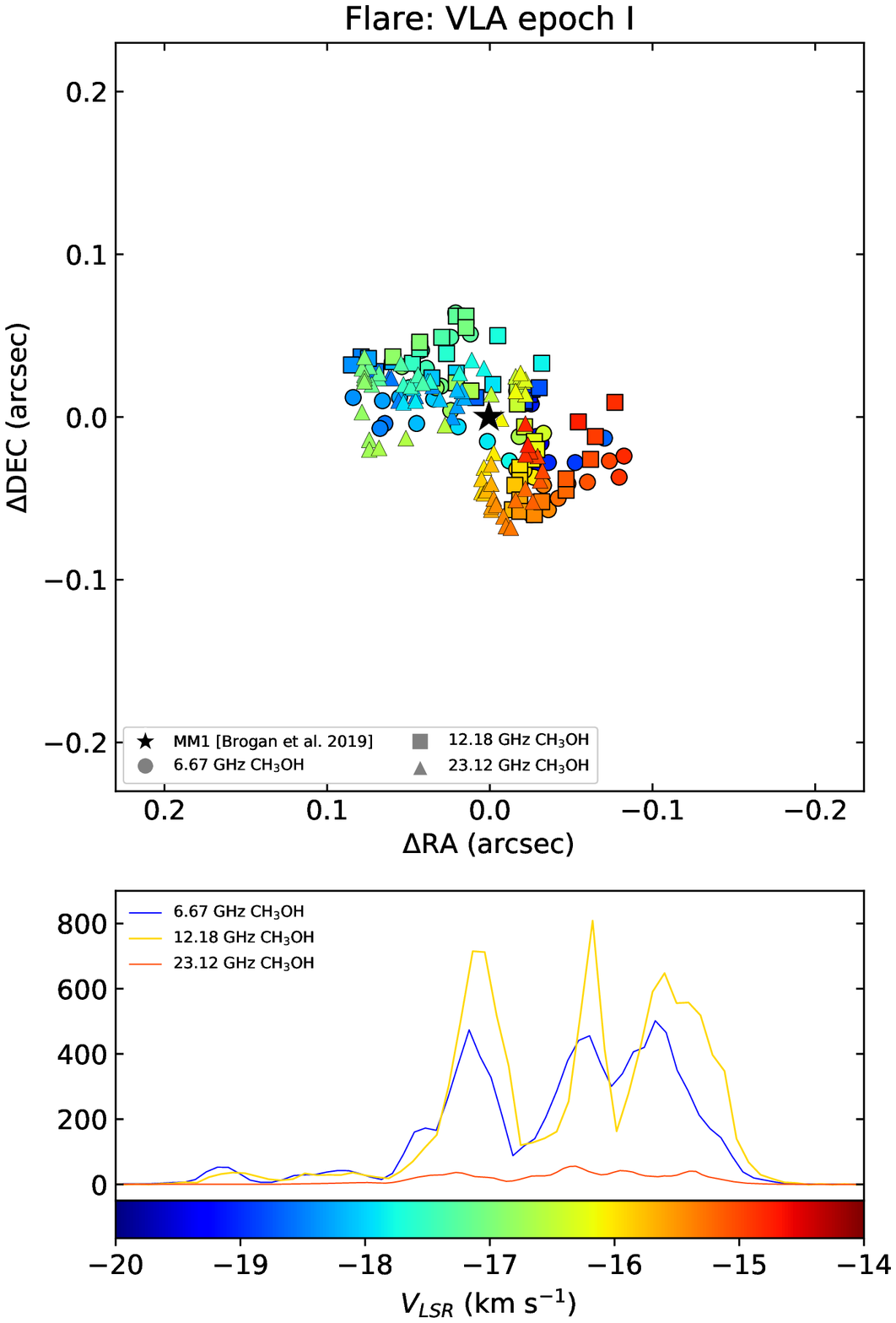}{0.45\textwidth}{(a)}
		  \fig{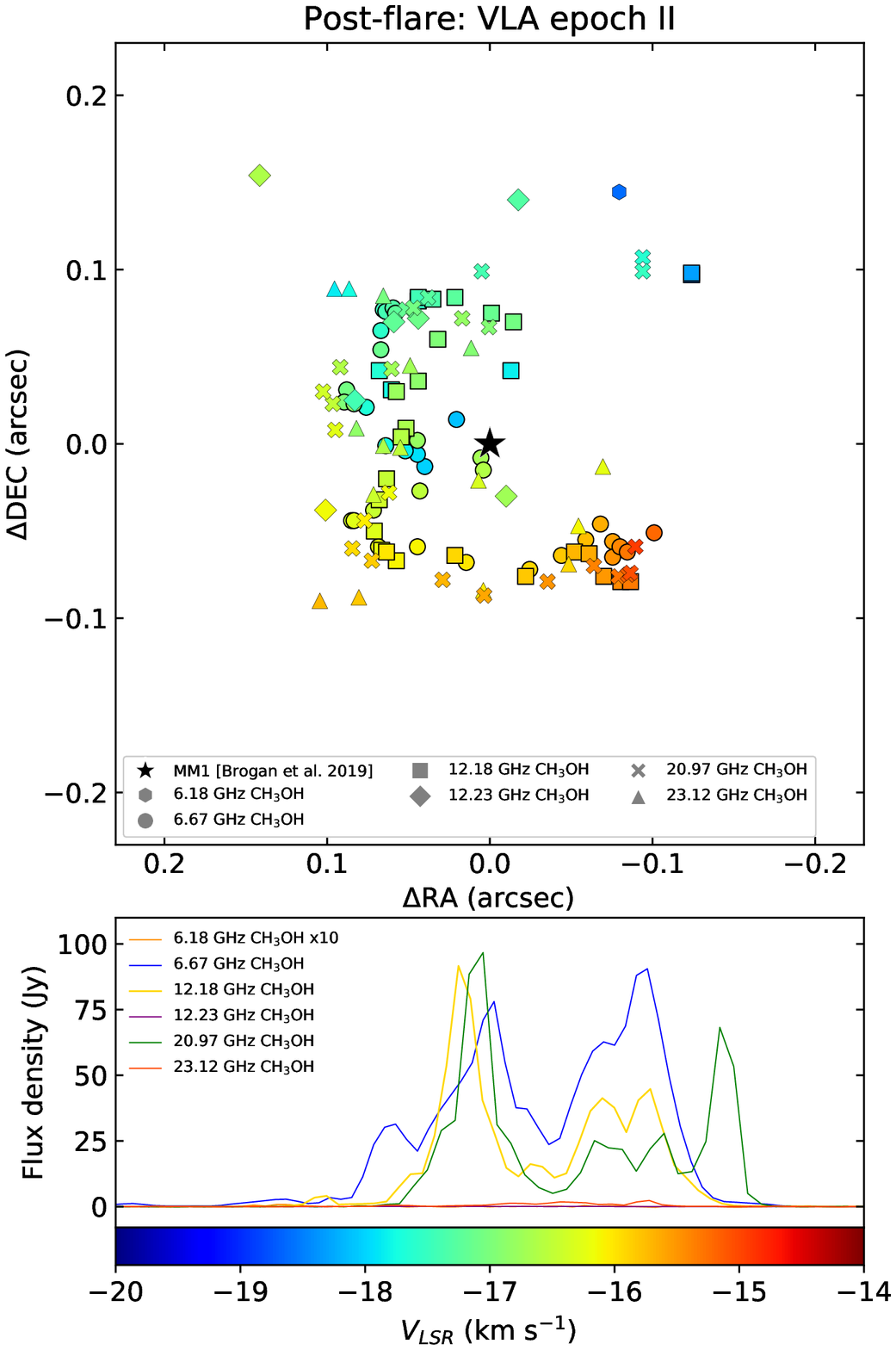}{0.44\textwidth}{(b)}
				  }
\caption{Comparison of  spatial distribution and spectra of the methanol 
maser emission detected in G358.93$-$0.03 at the (a) flare (VLA epoch I) and  (b) post-flare (VLA epoch II) epochs.
\textit{Top panels:} combined maser spot maps. Plots are color-coded by radial velocity (see colorbar for color scale).
\textit{Bottom panels:} spectra of the maser emission. \label{fig:compare2}}
\end{figure*}

Several of the newly detected methanol masers were imaged for the first time in the VLA observations presented here. 
The high sensitivity and moderate resolution of the VLA allow us to study the spatial structure of very weak masers, that are not accessible to VLBI imaging, but are readily detectable by the VLA. We were able to image and precisely locate positions of faint masers at 6.18 and 12.23 GHz with flux densities of $\sim$0.1~Jy (Figures \ref{fig:61GHZ} and~\ref{fig:122GHZ}).
Notably, not only do the imaged methanol masers, including the newly discovered ones, arise in the same region, but they also trace the same structures (Figure \ref{fig:compare2}).

The VLA epoch I (February 25, 2019) observations were carried out during the maximum flare activity of the 6.67 GHz methanol maser in G358.93$-$0.03, midway between the two peaks of the flare on February 14 and March 12 (iMet data; see section \ref{sec:results}). Thus, the  VLA epoch I can be considered as a flare epoch while the VLA epoch II (June 4, 2019) is a post-flare one.

Pre-flare VLA data from early 2012 exist and were published by \cite{Hu16}. In the quiescent state, the peak 6.67 GHz maser flux density was $\sim$5~Jy 
\citep{Hu16}. 
The data of \cite{Hu16} were obtained with the $\sim 4''$ beam of the VLA C-configuration, i.e. with much lower resolution than our VLA data, which precludes a comparison of the spot maps.

\begin{figure*}[ht]
\gridline{
\fig{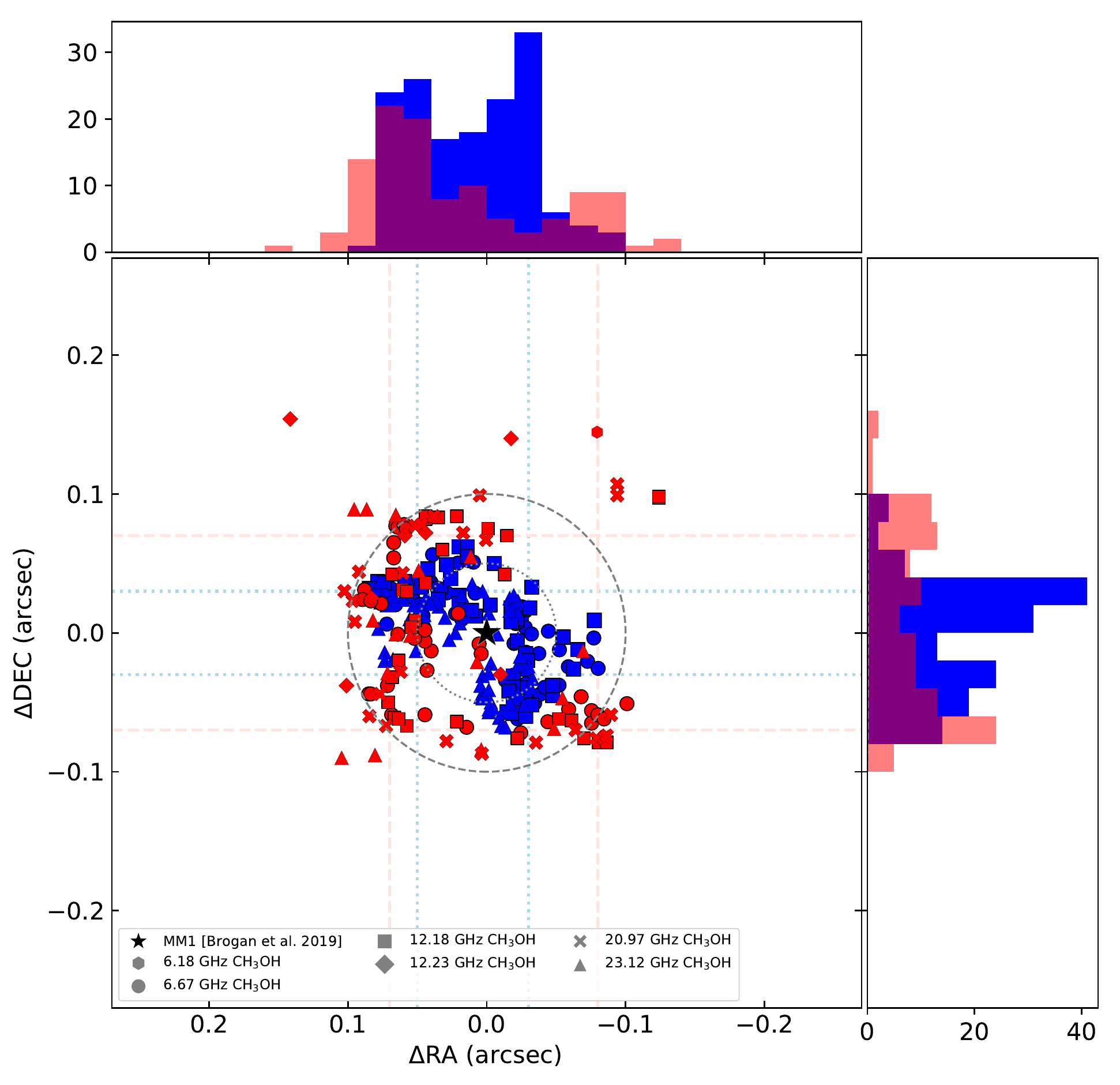}{0.7\textwidth}{}                          	  }
\caption{Overplot of the methanol maser features detected in G358.93$-$0.03 at the VLA epoch I  (blue markers) and II (red markers). The histograms show the number of the detected maser spots relative to right ascension and declination at the VLA epoch I  (blue) and II (red). The histogram bin size is 0.02~arcsec. The dotted blue and dashed red lines correspond to the maximums of the maser spot density at VLA epoch I and II, correspondingly. The dotted circle marks the distance at which the maximum density of the methanol maser emission was found at epoch I, and the dashed circle marks the distance for epoch II.  \label{fig:overplot}}
\end{figure*}

Our VLA data indicate that all of the flaring methanol masers originate from a single region of size $\sim$0.2$^{\prime\prime}$ (Figures \ref{fig:compare2} and \ref{fig:overplot}) which coincides with the brightest mm source, MM1, detected with ALMA \citep{Brogan19}. 
The close spatial association of the various flaring maser transitions is expected if caused by an accretion burst.

In order to analyse the maser density distribution at two VLA epochs and its evolution between the observations, we calculated the number of detected maser spots around MM1 -- see histograms in Figure \ref{fig:overplot}. The blue markers on the map and blue histogram bars correspond to the VLA epoch I, and the red markers/bars correspond to the epoch II. The bin width of the histograms is 0.02~arcsec which is one tenth of the area occupied by the maser emission and corresponds to the median position fitting error of our data (see Tables \ref{tab:T61GHZ}-\ref{tab:T23GHZ}).

At the first VLA epoch, the flaring methanol maser emission consists of two linked sub-clusters, where the northern cluster hosts blue-shifted velocity spots while the southern cluster has red-shifted ones (Figure \ref{fig:compare2}a). 
The analyses of the density of the maser spot distribution around MM1 (Figure \ref{fig:overplot}), shows two density peaks located almost symmetrically around the source at the separations of $\sim$0.05~arcsec. One 'knot' is located to NE and another to SW of the MM1 position.
Following the idea of \cite{Chen20b}, we interpret the VLA maser spot map obtained during the flare epoch for the 6.67, 12.18 and 23.12 class II methanol masers as a two-arm spiral pattern.
The most crowded parts of the maser clusters could be explained as turning points of the gas motion in the arms \citep{Meyer18}. 

At the post-flare epoch, all detected methanol masers (including the newly-discovered and rare transitions) trace a bow-shaped structure extending eastward from the MM1 position (Figure \ref{fig:compare2}b). While the velocity pattern persisted (the blue-shifted velocity spots are found to the north and the red-shifted to the south), the region affected by the maser emission expanded. 
The histograms in Figure~\ref{fig:overplot} show that the maximums of the maser spot density shifted outwards to the radius of  $\sim$0.09~arcsec from the position of the central source.

To estimate the evolution of the maser distribution in the region, we compare the positions of the maser emission density "knots" evaluated from the histograms in Figure~\ref{fig:overplot}. 
The density peaks were found at the diameter of $\sim$0.1~arcsec around MM1 at epoch I, and of $\sim$0.2~arcsec at epoch II. Thus, the region of physical conditions suitable to sustain maser emission doubled in size.

The drastic change between flare and post-flare epochs affected the entire masing region of $\sim$1000 AU (adopting the BeSSeL estimate of a near kinematic distance of 6.75 kpc) and happened in a period of about three months. Assuming these parameters, the triggering event may have propagated at a speed of $\sim$0.06$c$ from the bursting source.  Such a high speed is almost certainly not caused by any physical movement of material. Hence, the flare and spatial rearrangement of the maser emission must be caused by a radiative event from the protostar that propagated through regions of more favorable physical conditions, more distant from the protostar.

\subsection{Comparison of the VLA data with other observations} \label{sec:other}

A thermal radiation ``heatwave'' emanating from an accreting high-mass protostar in G358.93$-$0.03, propagating at subluminal speed, has been inferred from multi-epoch LBA observations of 6.67 GHz methanol masers \citep{Burns20}. 
The lower spatial resolution of the VLA prohibits such a precise analysis of the spatial structure of the maser emission in the region. On the other hand, it provides a unique insight into an extended component of maser emission. 

\begin{figure*}[ht]
\gridline{\fig{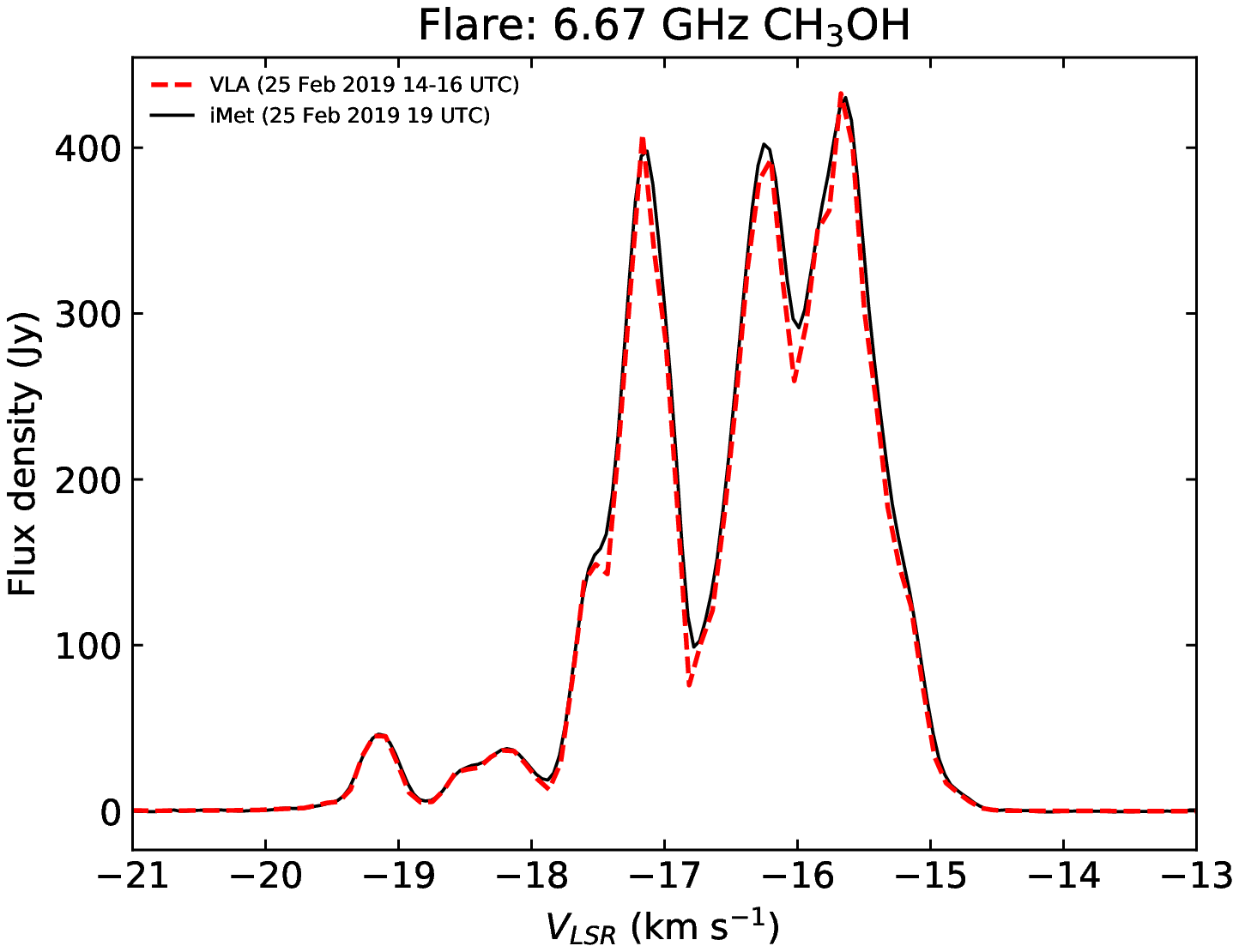}{0.48\textwidth}{(a)}
		  \fig{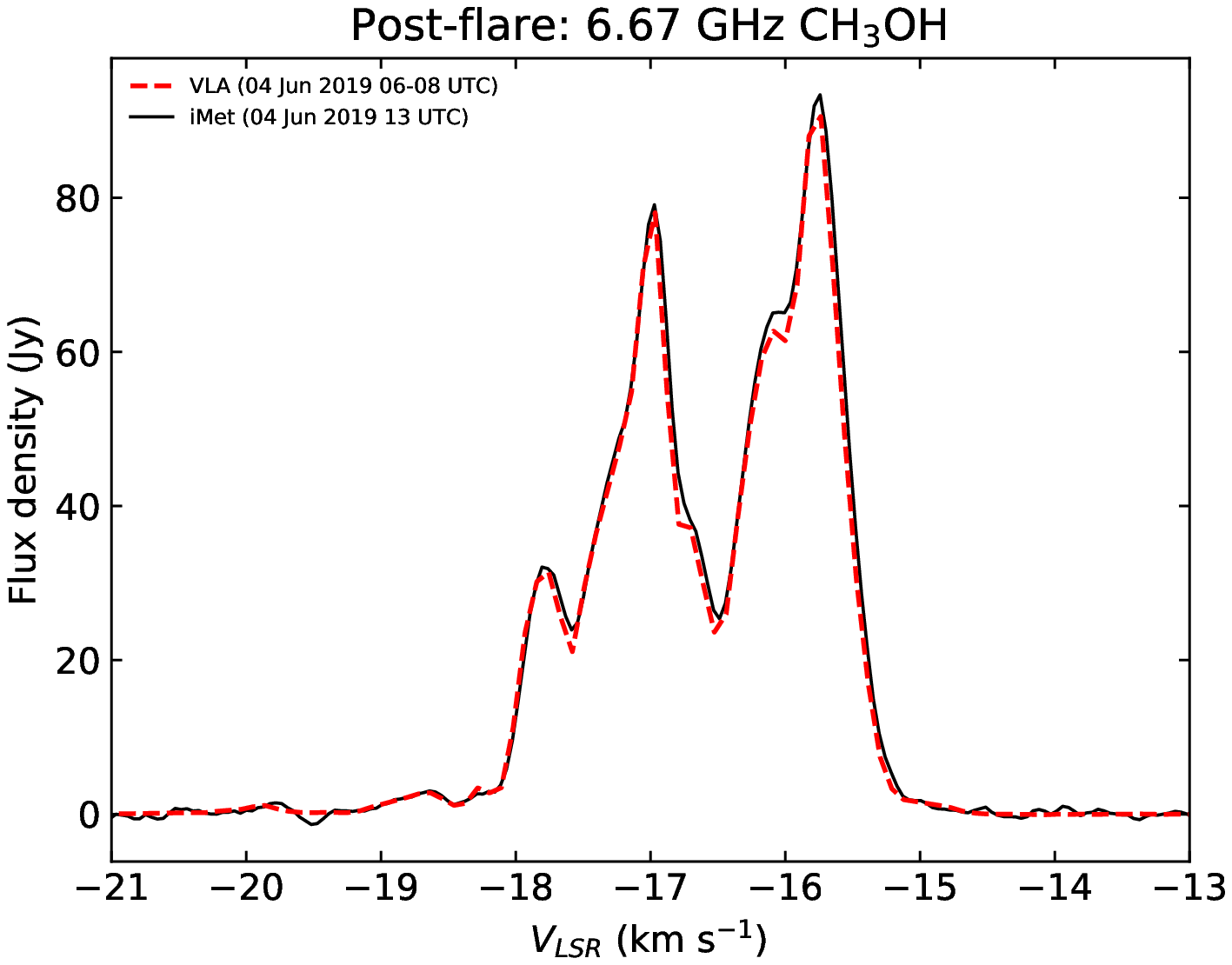}{0.48\textwidth}{(b)}
				  }
\caption{Comparison of the single-dish and cross-correlation VLA spectra of the 6.7 GHz methanol maser emission detected in G358.93$-$0.03 (a) on February 25, 2019 (epoch I) and (b) on June 4, 2019 (epoch II). The single-dish data is obtained in the the Ibaraki 6.7 GHz Methanol Maser Monitor (iMet) program. \label{fig:compspec}}
\end{figure*}

\begin{figure*}[ht]
\gridline{
\fig{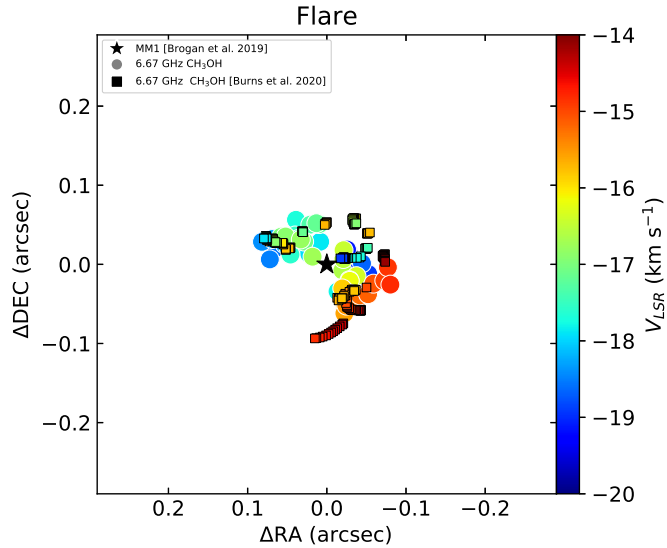}{0.5\textwidth}{}                          	  }
\caption{Comparison of the spatial distribution of the methanol maser emission detected in G358.93$-$0.03 with (a) the VLA on February 25, 2019 (this paper, epoch I) and LBA on February 28, 2019 \citep{Burns20}. \label{fig:lba}}
\end{figure*}

The comparison of the VLA cross-correlation spectra (with a synthesized beam size of $\sim$3.5 arcsec, see Table \ref{tab:obsspl}) and 
single-dish spectra obtained with the Hitachi 32-m radio telescope as a part of the Ibaraki 6.7-GHz Methanol Maser Monitor (iMet) program
shows that 100\% of the single-dish flux density is recovered on the VLA baselines (Figure \ref{fig:compspec}). In contrast, only $\sim$10\% of the maser flux density was recovered with the milli-arcsecond synthesized beam of the LBA \citep{Burns20}.
Based on these results and in accordance with the core/halo hypothesis of the structure of class II methanol maser masers (e.g. \citealt{Minier2002}), we conclude that with the VLA baselines we detect both compact (core) and extended (halo) emission while with the LBA baselines we detect only the compact core emission.

The comparison of the 6.7 GHz methanol maser spot maps obtained in the VLBI observations of \cite{Burns20} and in our VLA observations (Figure~\ref{fig:lba}), shows that compact and extended components of the maser emission trace the same region around MM1, but highlight rather different structures. Both compact (VLBI data) and extended (VLA data) emission can be divided in two clusters: an upper (northern) cluster consisting of blue-shifted velocity spots and a lower (southern) cluster consisting of red-shifted velocity spots. However, the northern cluster of extended VLA maser emission is smaller than the southern cluster, while for the compact VLBI emission, the two  clusters have about the same size. Also noteworthy is that both extended and compact maser emission trace roughly the same turning points in the both the northern and southern clusters.

Nevertheless, the LBA \citep{Burns20} and VLA  observations were carried out with a three-day interval from each other during the period of high activity of the source. Therefore, it is possible that not all differences in the structure of the spot maps can be attributed to the extended/compact emission duality. For example, the most southern, linearly elongated maser  cluster detected with the LBA has no counterpart in the VLA map. The iMet monitoring data indicated that the source shows daily/intraday variability.
If we assume that a $\sim$0.05 arcsec cluster was ignited by the propagation of the heat wave over three days between the observations, the wave speed must be of $\sim$0.7c. 
On the one hand, a wave propagating through the low density material is supposed to have a speed of light, thus the speed of $\sim$0.7c is achievable, especially in the lower density outer regions. 
However, the absence of the elongated southern substructure is most likely due to the effect of centroid mapping. The most distant spot from MM1 with a particular velocity will be averaged in the synthesised beam with any other spots with the same velocity in the region, on average that will bring the centroid map position to be closer to center of the red cluster.

\begin{figure*}[ht]
\gridline{\fig{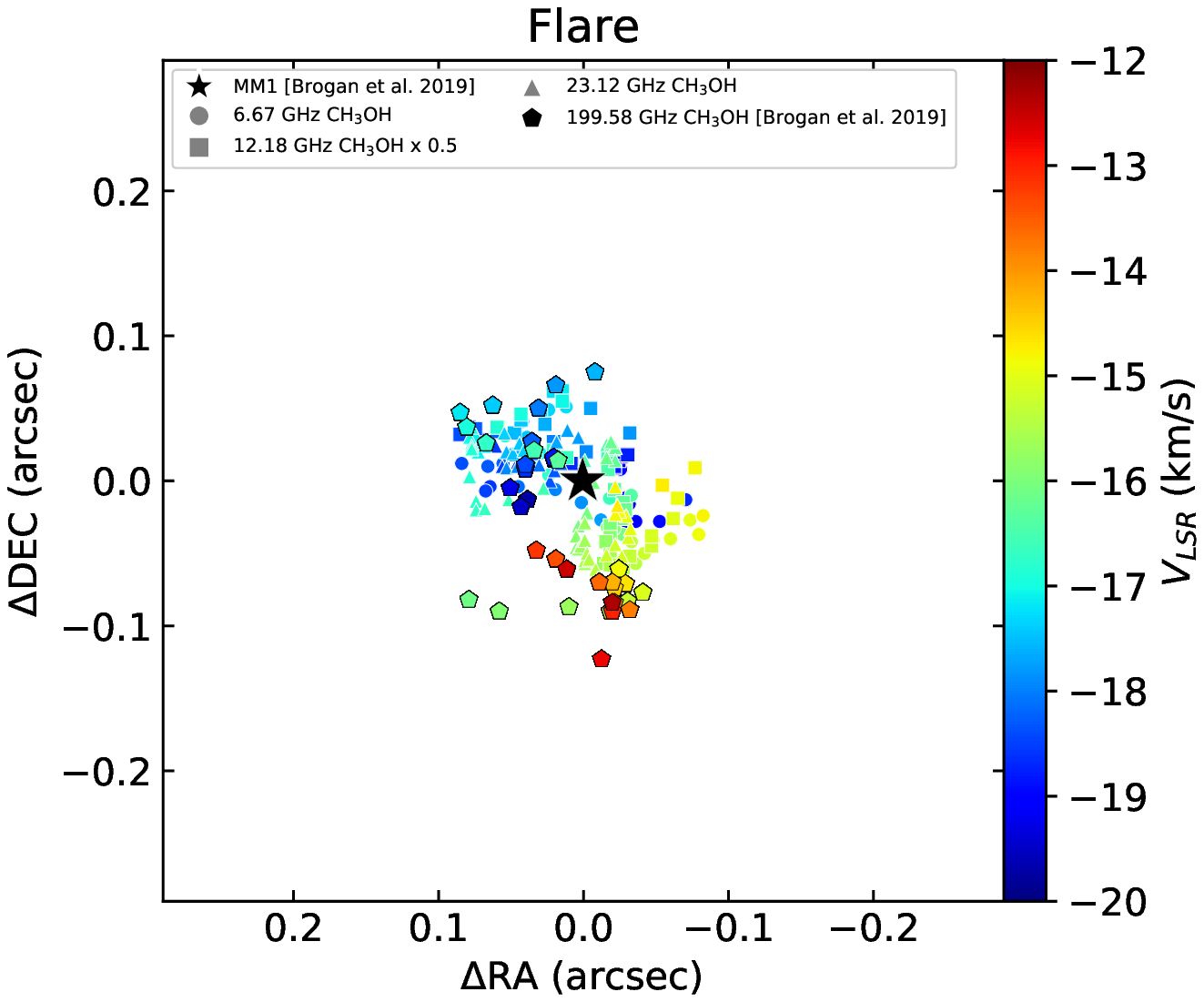}{0.48\textwidth}{(a)}
		  \fig{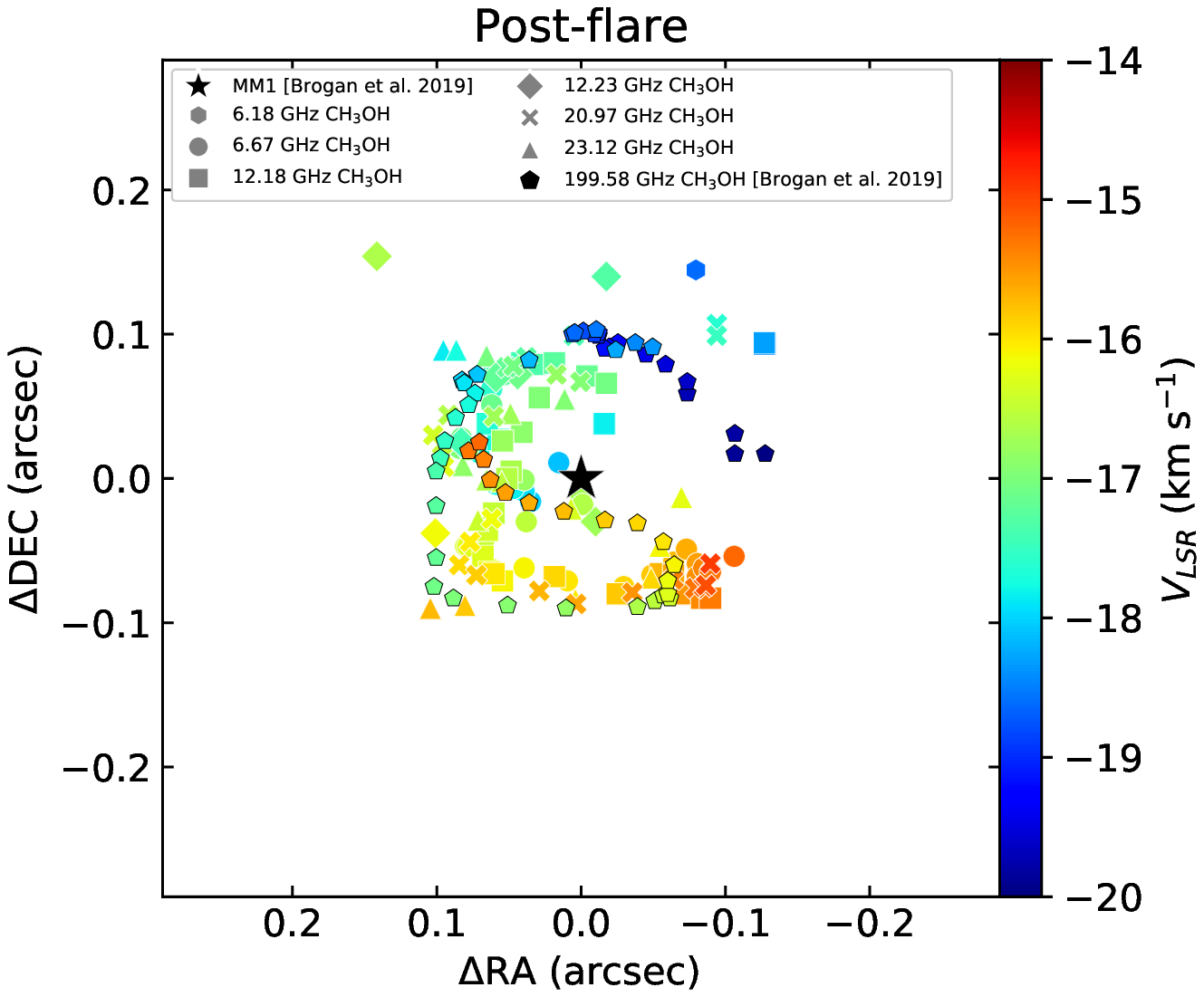}{0.48\textwidth}{(b)}
				  }
\caption{Comparison of spatial distribution of the methanol maser emission detected in G358.93$-$0.03 with  with (a) the VLA on February 25, 2019 (this paper, epoch I) and SMA on March 14, 2019 \citep{Brogan19}; (b) the VLA on June 4, 2019 (this paper, epoch II) and ALMA on April 16, 2019 \citep{Brogan19}. \label{fig:alma}}
\end{figure*}

\begin{figure*}[ht]
\gridline{\fig{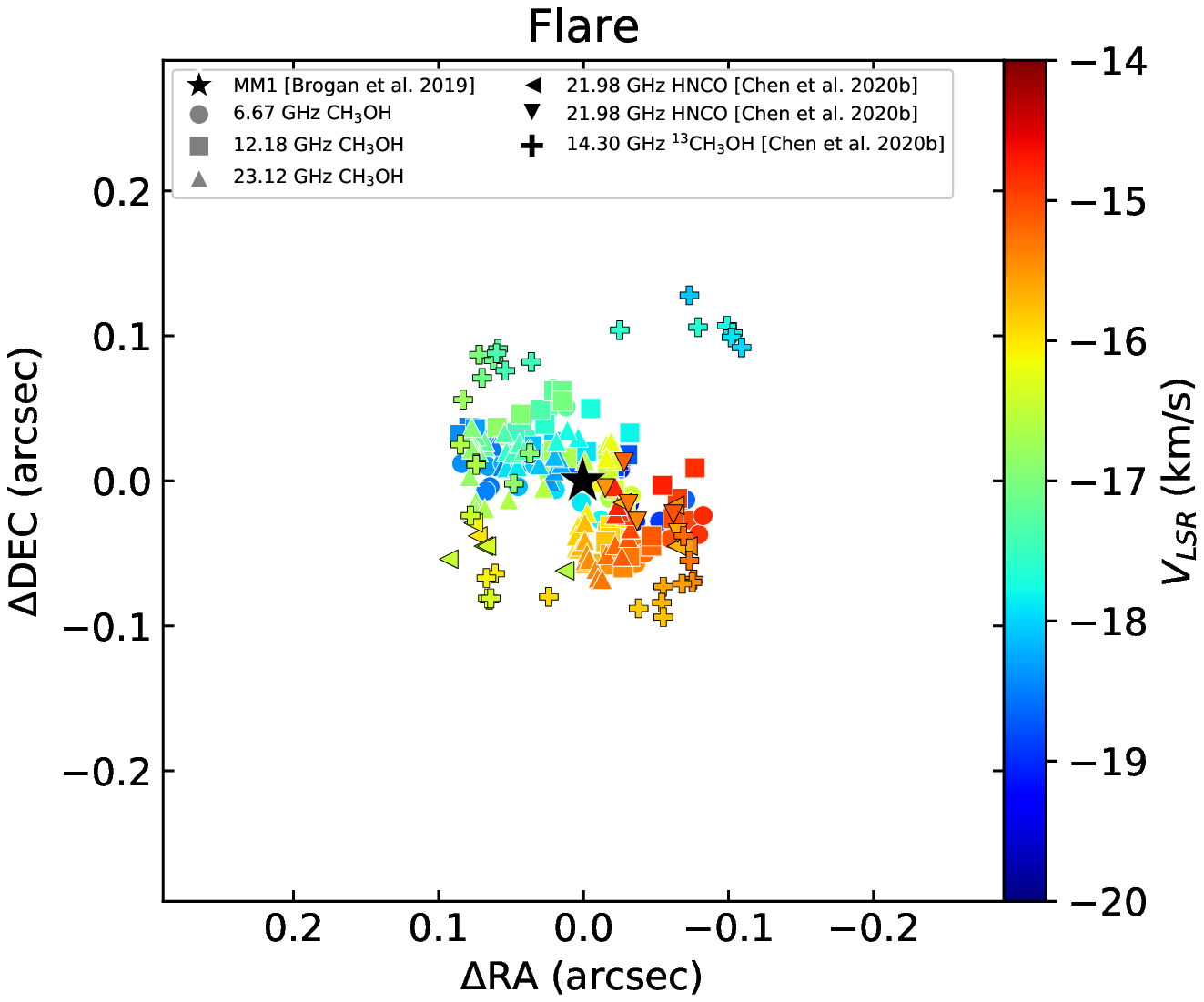}{0.48\textwidth}{(a)}
		  \fig{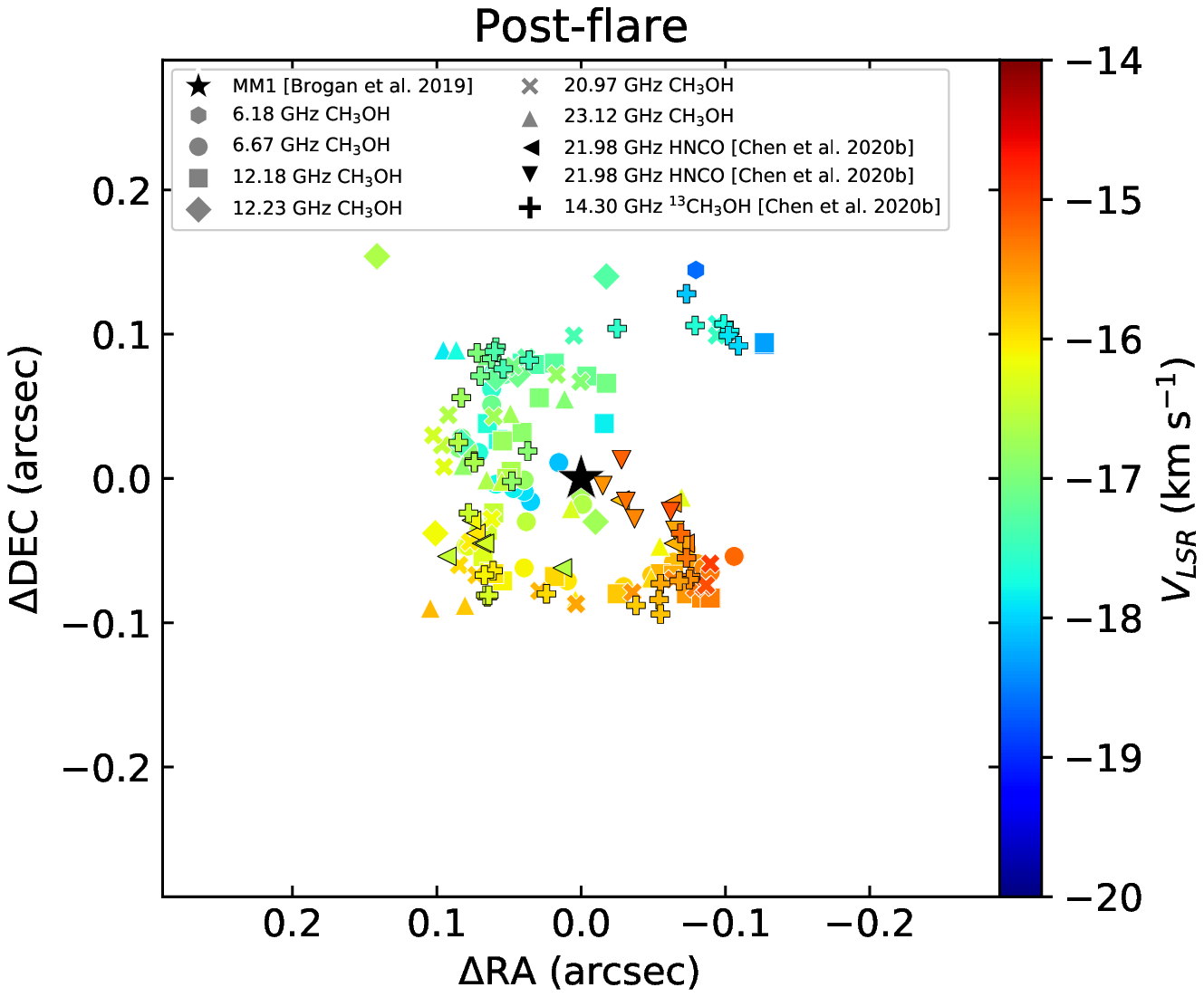}{0.48\textwidth}{(b)}
				  }
\caption{Comparison of spatial distribution of the methanol maser emission detected in G358.93$-$0.03 with the VLA on March 24, 2019 \citep{Chen20b} and (a) the VLA on February 25, 2019 (this paper, epoch I), (b) the VLA on June 4, 2019 (this paper, epoch II). \label{fig:comparechen}}
\end{figure*}

It is also found that the cm-wavelength methanol masers detected with the VLA (this paper) and newly discovered high-frequency methanol masers detected with the SMA and ALMA \citep{Brogan19} trace similar spatial patterns (see Figure \ref{fig:alma}). Here we present a comparison of our VLA data with the 199.575 GHz methanol maser -- the strongest (sub)millimeter maser detected in \cite{Brogan19}. Our analysis showed that the VLA epoch I data (February, 2019) resembles the SMA 199.575 GHz maser data obtained on March 14, 2019, while the VLA epoch II data (June, 2019) is similar to the spatial distribution of the 199.575 GHz methanol maser detected on April, 16 with the ALMA. In February-March, 2019, during the peak activity of the methanol maser flare (see Figure \ref{fig:timeline}), both the common (VLA data) and (sub)millimeter, torsionally excited (SMA and ALMA data) methanol maser transitions are found  in two sub-clusters separated by location and velocity, with the northern cluster containing the blue-shifted maser spots and the southern cluster predominated by the red-shifted velocity spots (Figure \ref{fig:alma}a). At the post-flare epoch, in April-June, 2019, all detected methanol masers trace a wider and less-dense/more-elongated formation, which can be interpreted as a spiral-arm structure (Figure \ref{fig:alma}b).

The cross-matching of our data and the results from  \cite{Chen20b} showed that the spatial distribution of the methanol masers detected at the second VLA epoch closely reassembles that which \cite{Chen20b} reported for the $^{13}$CH$_3$OH, HDO, and HNCO masers in G358.93$-$0.03 (Figure~\ref{fig:comparechen}b). 
Note that despite the fact that the observations of \cite{Chen20b} were conducted with the VLA on April 4, 2019, i.e. closer to our VLA epoch I (38 day apart - see Figure~\ref{fig:comparechen}a), the data of the VLA epoch II (61 day apart) appeared to be more fitting (Figure~\ref{fig:comparechen}b).
\cite{Chen20b} argue that the rare masers detected in their VLA observation trace a two-armed spiral structure; in this scenario the arms reveal  accretion flows falling onto the central star in spiral trajectories. According to theory (e.g. \citealt{Meyer2019, Jankovic19}), spiral arms, as well as accretion burst, are products of gravitational instability of a disk around MYSOs. 
Our VLA data seems to suggest that the spiral-arm features of the accretion disk are traced not only by rare maser species detected by \cite{Chen20b}, but also by more common methanol masers at 6.67 and 12.18 GHz.
Note that the observations of \cite{Chen20b} were performed still at the flare epoch (April 4), while our VLA observations took place at the post-flare epoch (June 4) - see Figure \ref{fig:timeline}. In the period between observations, no drastic changes in the structure of the spectrum were observed for the methanol maser at 6.7 GHz, for example; the only striking difference was the gradual decrease in the maser flux density from $\sim$500 Jy to $\sim$100 Jy\footnote{\url{http://vlbi.sci.ibaraki.ac.jp/iMet/G358.9-00-190114/daily.html}}. 
The two-arm spiral infall gas flow structure seems to still exist at the post-flare period, and the physical conditions (e.g., temperature) in these flows change with the flare propagation, making different type of masers (with different pumping conditions) gradually arise or disappear in the flows.
The resemblance of the maser clusters detected in \cite{Chen20b} and in our observations at epoch II (Figure \ref{fig:comparechen}b) indicates that the spatial distribution of the methanol masers remained roughly constant after April 2019.

Note that the most recent VLBI/ALMA results obtained by the M2O indicate the presence of a more complex spatial structure of the regions traced by the methanol masers than that found with the VLA (Burns et al. in prep., Brogan et al. in prep).
For example, VLBI observations suggest a 4-arm system (Burns et al. in prep.), not the two-arm system that was discovered with the VLA (\cite{Chen20b}, this paper). Nevertheless, the VLBI and VLA results do not contradict each other, but rather indicate the limitations of spatially unresolved maser spot
fitting when there is complex underlying spatial morphology in each
channel.
It is possible that the two arms in the blue side of the disk are averaged within the VLA synthesised beam into one arm, and the same for the two arms on the red side of the disk. Since the disk is almost face-on, the velocity differences across the disk are small which worsens this effect as the centroid mapping technique derives a single spot per velocity channel and many maser spots exist in different parts of the disk (in different arms) at the same velocity. A more detailed comparison of the VLA and VLBI/ALMA results obtained for the source will be presented in the future M2O publications.

\subsection{Cm-continuum emission of MM1}

The strong millimeter continuum emission of MM1 \cite{Brogan19}, coupled with the lack of significant centimeter emission (Figure \ref{fig:cont}) indicates that the source is probably in a precursor state of ultracompact HII region \citep{Churchwell02}, as it was indicated in \cite{Brogan19}.

Our continuum observations suggest a moderate decrease in the flux densities of MM1 in the period between the flare and post-flare epochs. 
As we indicated in Section \ref{sec:results}, the ALMA mm-observations and VLA cm-observations were performed during different epochs of the source activity.
The analysis is particularly difficult because the presumptive bursting source MM1 was detected only in K-band and at one epoch of observation.  Moreover, in contrast to  S255 and NGC6334I, cm-continuum emission in G358.93$-$0.03 shows much lower flux densities. 

Nevertheless, a decrease of the cm-continuum flux density of a source can happen during accretion event.
In the case of NGC6334I, for example, \cite{Hunter17} reported a 4-fold increase in the dust continuum emission, while the free-free emission (at 1.5 cm) fell by about the same factor \citep{Brogan2018, Hunter2019}.
Such a decrease would naturally occur if the ionizing photon flux from the young stellar object is reduced due to the bloating of the star.
At the high densities of these MSFR, the recombination timescale for the ionized gas can be of the order of days to weeks. Hence, the free-free continuum emission can track the same accretion events, but show a decrease while the IR luminosity increases \citep{Hunter2019}.

\section{Summary} \label{sec:conclusions}

A multi-epoch and multi-frequency VLA imaging of maser and continuum  emission in C, Ku, and K-bands was performed for the  massive  star-forming  region G358.93$-$0.03. Two observing sessions were conducted, at the maser flare epoch (VLA epoch I) and at a post-flare epoch (VLA epoch II). The main outcomes and scientific 
insights obtained from the observations are summarized as follows:

(1) Maser emission is detected and imaged in  
several methanol transitions. 
Spatial structure evolution is studied  for methanol masers at 6.67, 12.18, and 23.12 GHz
at two observational epochs.

(2) The first interferometric images are obtained for the new  methanol  maser  transitions  at  6.18,  12.23,  and  20.97  GHz (the masers were discovered in  G358.93$-$0.03 in  single-dish  and  ATCA observations \citep{MacLeod19,Brogan19}).

(3) Methanol maser emission in all detected transitions and at both epochs is found in a region with a diameter of $\sim$0.2$^{\prime\prime}$ around the MM1 position and shows a velocity gradient in the NS direction.

(4) A drastic change in the spatial distribution of the detected methanol masers is found.
At the flare epoch, the 6.67, 12.18, and 23.12 GHz methanol masers are found in elongated regions, aligned in a NE-SW direction.
At the post-flare epoch, the methanol masers trace bow-shaped structures extending eastward from the MM1 position. During  the  transition  from  the  first  epoch  to  the  second,  the  region  traced  by  masers expanded while the velocity gradient decreased.

(5) The obtained data suggests that the methanol masers detected with the VLA trace  the spiral-arm structures within the accretion disk which were first discovered in rare maser lines in \cite{Chen20b}.

(6) K-band continuum emission is detected at VLA epoch I toward the supposed bursting source and the most line-rich hot core, detected with ALMA by \cite{Brogan19} as G358.93$-$0.03-MM1. A  moderate  decrease  in  the  flux  density  of  MM1  in  the  period between the flare and post-flare epoch is detected.

\newpage
\acknowledgments

The Ibaraki 6.7-GHz Methanol Maser Monitor (iMet) program is partially supported by the Inter-university collaborative project "Japanese VLBI Network (JVN)" of NAOJ and JSPS KAKENHI Grant Numbers JP24340034, JP21H01120, and JP21H00032 (YY). X. C. thanks to the supports from the
National Natural Science Foundation of China (11873002, 11590781, 11903009), and Guangdong Province Universities and Colleges Pearl River 
Scholar Funded Scheme (2019). 

A.M.S. was supported by the Foundation for the Advancement of Theoretical Physics and Mathematics “BASIS” for his work on section 4.1 and by the Russian Science Foundation grant No. 18-12-00193-P for his work on other sections.

The National Radio Astronomy Observatory is a facility of the National Science Foundation operated under cooperative agreement by Associated Universities, Inc. This paper makes use of the following ALMA data: ADS/JAO.ALMA\#2018.A.00031.TS.  ALMA is a partnership of ESO (representing its member states), NSF (USA) and NINS (Japan), together with NRC (Canada), MOST and ASIAA (Taiwan), and KASI (Republic of Korea), in cooperation with the Republic of Chile. The Joint ALMA Observatory is operated by ESO, AUI/NRAO and NAOJ.  The SMA is a joint project between the Smithsonian Astrophysical Observatory and the Academia Sinica Institute of Astronomy and Astrophysics and is funded by the Smithsonian Institution and the Academia Sinica.

\clearpage
\appendix

\startlongtable
\begin{deluxetable}{cccccc}
\tablewidth{0pt}
\tablecaption{6.18 GHz CH$_3$OH maser parameters \label{tab:T61GHZ}}
\tablehead{
\colhead{Epoch} &
\colhead{RA(J2000)} & \colhead{DEC(J2000)} &
\colhead{Integrated flux} &
\colhead{Peak flux} &
\colhead{V$_{LSR}$} \\
\colhead{} & 
\colhead{($^h$~$^m$~$^s$)} & \colhead{($^\circ$~$\arcmin$~$\arcsec$)} &
\colhead{(Jy)} & 
 \colhead{(Jy/beam)} & \colhead{(km~s$^{-1}$)} 
}
\startdata
II & 17:43:10.0961$\pm$0.0012 & -29:51:45.549$\pm$0.106  & 0.08$\pm$0.01  & 0.09$\pm$0.01 & -18.61  \\ 
\enddata
\end{deluxetable}

\startlongtable
\begin{deluxetable}{cccccc}
\tablewidth{0pt}
\tablecaption{6.67 GHz CH$_3$OH maser parameters \label{tab:T67GHZ}}
\tablehead{
\colhead{Epoch} &
\colhead{RA(J2000)} & \colhead{DEC(J2000)} &
\colhead{Integrated flux} &
\colhead{Peak flux} &
\colhead{V$_{LSR}$} \\
\colhead{} & 
\colhead{($^h$~$^m$~$^s$)} & \colhead{($^\circ$~$\arcmin$~$\arcsec$)} &
\colhead{(Jy)} & 
 \colhead{(Jy/beam)} & \colhead{(km~s$^{-1}$)} 
}
\startdata
I & 17:43:10.0989$\pm$0.0001 & -29:51:45.757$\pm$0.014 & 15.55$\pm$0.15 & 15.80$\pm$0.06 & -19.36\\
& 17:43:10.0988$\pm$0.0001 & -29:51:45.786$\pm$0.010 & 38.08$\pm$0.27 & 38.12$\pm$0.11 & -19.27\\
& 17:43:10.0989$\pm$0.0001 & -29:51:45.768$\pm$0.009 & 54.48$\pm$0.34 & 54.16$\pm$0.13 & -19.19\\
\enddata
\tablecomments{(1) Table \ref{tab:T67GHZ} is published in its entirety in the machine-readable format. A portion is shown here for guidance regarding its form and content. \\ (2) The positional shifts of $\Delta$RA=0.010, $\Delta$DEC=0.079 (epoch I) and $\Delta$RA=0.017, $\Delta$DEC=0.017 (epoch II) were introduced to the data to prepare Figures \ref{fig:67GHZ} and \ref{fig:compare2}-\ref{fig:overplot}, and \ref{fig:lba}-\ref{fig:comparechen} (see section \ref{sec:obs}).} 
\end{deluxetable}

\startlongtable
\begin{deluxetable}{cccccc}
\tablewidth{0pt}
\tablecaption{12.18 GHz CH$_3$OH maser parameters \label{tab:T12GHZ}}
\tablehead{
\colhead{Epoch} &
\colhead{RA(J2000)} & \colhead{DEC(J2000)} &
\colhead{Integrated flux} &
\colhead{Peak flux} &
\colhead{V$_{LSR}$} \\
\colhead{} & 
\colhead{($^h$~$^m$~$^s$)} & \colhead{($^\circ$~$\arcmin$~$\arcsec$)} &
\colhead{(Jy)} & 
 \colhead{(Jy/beam)} & \colhead{(km~s$^{-1}$)} 
}
\startdata
I & 17:43:10.1024$\pm$0.0100 & -29:51:45.619$\pm$0.007 & 21.70$\pm$0.21 & 21.95$\pm$0.09 & -19.25 \\
  & 17:43:10.1020$\pm$0.0100 & -29:51:45.621$\pm$0.007 & 31.13$\pm$0.29 & 31.30$\pm$0.12 & -19.16 \\
  & 17:43:10.1022$\pm$0.0100 & -29:51:45.626$\pm$0.007 & 35.83$\pm$0.33 & 35.92$\pm$0.14 & -19.06 \\
\enddata
\tablecomments{(1) Table \ref{tab:T12GHZ} is published in its entirety in the machine-readable format. A portion is shown here for guidance regarding its form and content. \\ (2) The positional shifts of $\Delta$RA=-0.034, $\Delta$DEC=-0.058 (epoch I) and $\Delta$RA=-0.070, $\Delta$DEC=-0.058 (epoch II) were introduced to the data to prepare Figures \ref{fig:12GHZ} and \ref{fig:compare2}-\ref{fig:overplot}, and \ref{fig:lba}-\ref{fig:comparechen}(see section \ref{sec:obs}).}  
\end{deluxetable}

\newpage
\startlongtable
\begin{deluxetable}{cccccc}
\tablewidth{0pt}
\tablecaption{12.23 GHz CH$_3$OH maser parameters \label{tab:T122GHZ}}
\tablehead{
\colhead{Epoch} &
\colhead{RA(J2000)} & \colhead{DEC(J2000)} &
\colhead{Integrated flux} &
\colhead{Peak flux} &
\colhead{V$_{LSR}$} \\
\colhead{} & 
\colhead{($^h$~$^m$~$^s$)} & \colhead{($^\circ$~$\arcmin$~$\arcsec$)} &
\colhead{(Jy)} & 
 \colhead{(Jy/beam)} & \colhead{(km~s$^{-1}$)} 
}
\startdata
II & 17:43:10.1148$\pm$0.0012 & -29 51 45.715$\pm$0.080 & 0.09$\pm$0.02 & 0.07$\pm$0.01 &-16.15\\
   & 17:43:10.1175$\pm$0.0008 & -29 51 45.523$\pm$0.006 & 0.08$\pm$0.02 & 0.08$\pm$0.01 &-16.63\\
   & 17:43:10.1074$\pm$0.0013 & -29 51 45.707$\pm$0.057 & 0.10$\pm$0.02 & 0.08$\pm$0.01 &-16.73\\
\enddata
\tablecomments{(1) Table \ref{tab:T122GHZ} is published in its entirety in the machine-readable format. A portion is shown here for guidance regarding its form and content. \\ (2) The positional shift of $\Delta$RA=-0.1, $\Delta$DEC=-0.016 (epoch II) was introduced to the data to prepare Figures \ref{fig:122GHZ} and \ref{fig:compare2}-\ref{fig:overplot}, and \ref{fig:lba}-\ref{fig:comparechen} (see section \ref{sec:obs}).} 
\end{deluxetable}

\startlongtable
\begin{deluxetable}{cccccc}
\tablewidth{0pt}
\tablecaption{20.97 GHz CH$_3$OH maser parameters \label{tab:T20GHZ}}
\tablehead{
\colhead{Epoch} &
\colhead{RA(J2000)} & \colhead{DEC(J2000)} &
\colhead{Integrated flux} &
\colhead{Peak flux} &
\colhead{V$_{LSR}$} \\
\colhead{} & 
\colhead{($^h$~$^m$~$^s$)} & \colhead{($^\circ$~$\arcmin$~$\arcsec$)} &
\colhead{(Jy)} & 
 \colhead{(Jy/beam)} & \colhead{(km~s$^{-1}$)} 
}
\startdata
II &17:43:10.0988$\pm$0.0003 &-29:51:45.608$\pm$0.012 & 7.31$\pm$0.36 & 5.95$\pm$0.16 & -17.61 \\
   &17:43:10.0988$\pm$0.0003 &-29:51:45.616$\pm$0.013 &13.29$\pm$0.64 &10.31$\pm$0.28 & -17.50 \\
   &17:43:10.1053$\pm$0.0003 &-29:51:45.616$\pm$0.012 &27.30$\pm$1.20 &18.48$\pm$0.48 & -17.39 \\
\enddata
\tablecomments{(1) Table \ref{tab:T20GHZ} is published in its entirety in the machine-readable format. A portion is shown here for guidance regarding its form and content. \\ (2) The positional shift of $\Delta$RA=-0.055, $\Delta$DEC=0.022 (epoch II) was introduced to the data to prepare Figures \ref{fig:20GHZ} and \ref{fig:compare2}-\ref{fig:overplot}, and \ref{fig:lba}-\ref{fig:comparechen} (see section \ref{sec:obs}).} 
\end{deluxetable}

\startlongtable
\begin{deluxetable}{cccccc}
\tablewidth{0pt}
\tablecaption{23.12 GHz CH$_3$OH maser parameters \label{tab:T23GHZ}}
\tablehead{
\colhead{Epoch} &
\colhead{RA(J2000)} & \colhead{DEC(J2000)} &
\colhead{Integrated flux} &
\colhead{Peak flux} &
\colhead{V$_{LSR}$} \\
\colhead{} & 
\colhead{($^h$~$^m$~$^s$)} & \colhead{($^\circ$~$\arcmin$~$\arcsec$)} &
\colhead{(Jy)} & 
 \colhead{(Jy/beam)} & \colhead{(km~s$^{-1}$)} 
}
\startdata
I &17:43:10.1055$\pm$0.0001 &-29:51:45.720$\pm$0.014&  0.87$\pm$0.03&  0.82$\pm$0.01& -18.59 \\
  &17:43:10.1054$\pm$0.0001 &-29:51:45.720$\pm$0.012&  1.38$\pm$0.04&  1.19$\pm$0.02& -18.54 \\
  &17:43:10.1051$\pm$0.0001 &-29:51:45.734$\pm$0.010&  1.99$\pm$0.05&  1.63$\pm$0.02& -18.49 \\
\enddata
\tablecomments{(1) Table \ref{tab:T23GHZ} is published in its entirety in the machine-readable format. A portion is shown here for guidance regarding its form and content. \\ (2) The positional shifts of $\Delta$RA=-0.005, $\Delta$DEC=-0.051 (epoch I) and $\Delta$RA=-0.035, $\Delta$DEC=-0.040 (epoch II) were introduced to the data to prepare Figures \ref{fig:23GHZ} and \ref{fig:compare2}-\ref{fig:overplot}, and \ref{fig:lba}-\ref{fig:comparechen} (see section \ref{sec:obs}).} 
\end{deluxetable}

\clearpage
\bibliography{bib}{}
\bibliographystyle{aasjournal}





\end{document}